\documentclass[12pt,a4]{article}
\pdfoutput=1

\usepackage{latexsym,epsfig,amssymb,euscript,amsmath,verbatim,cite, subfigure,lmodern,color,verbatim}
\usepackage{float}
\usepackage{verbatim} 
\newcommand{\be}{\begin{equation}}
\newcommand{\ee}{\end{equation}}
\newcommand{\bea}{\begin{eqnarray}}
\newcommand{\eea}{\end{eqnarray}}

\numberwithin{equation}{section}
\usepackage{bm}
\usepackage{hyperref}

\begin{document}
\pagestyle{empty}
%\today
\vspace{1.8cm}
% \begin{flushright}
% {\small{
% Preprint}}
% \end{flushright}

\begin{center}
{\LARGE{\bf  {Thermo-electric transport in gauge/gravity models with momentum dissipation}}}

\vspace{1cm}

{\large{Andrea Amoretti$^{a,}$\footnote{\tt andrea.amoretti@ge.infn.it },
Alessandro Braggio$^{b,}$\footnote{\tt alessandro.braggio@spin.cnr.it }, 
Nicola Maggiore$^{a,}$\footnote{\tt nicola.maggiore@ge.infn.it },\\ 
Nicodemo Magnoli$^{a,}$\footnote{\tt nicodemo.magnoli@ge.infn.it},
Daniele Musso$^{c,}$\footnote{\tt dmusso@ictp.it} 
\\[1cm]}}

{\small{
{}$^a$  Dipartimento di Fisica, Universit\`a di Genova,\\
via Dodecaneso 33, I-16146, Genova, Italy\\and\\I.N.F.N. - Sezione di Genova\\
\medskip
{}$^b$  CNR-SPIN, Via Dodecaneso 33, 16146, Genova, Italy\\
\medskip
{}$^c$ Abdus Salam International Centre for Theoretical Physics (ICTP)\\
Strada Costiera 11, I 34014 Trieste, Italy
}}
\vspace{1cm}

{\bf Abstract}

We present a systematic definition and analysis of the thermo-electric linear response in gauge/gravity systems
focusing especially on models with massive gravity in the bulk and therefore momentum dissipation in the dual field theory.
A precise treatment of finite counter-terms proves to be essential to yield a consistent physical picture whose hydrodynamic and 
beyond-hydrodynamics behaviors noticeably match with field theoretical expectations. The model furnishes
a possible gauge/gravity description of the crossover from the quantum-critical to the disorder-dominated Fermi-liquid behaviors, as expected in graphene.

\end{center}

\newpage

%  Resetting of counters
\setcounter{page}{1} \pagestyle{plain} \renewcommand{\thefootnote}{\arabic{footnote}} \setcounter{footnote}{0}

\tableofcontents

\section{Introduction and motivation}

An accurate physical description of real condensed matter systems usually requires to include mechanisms 
for momentum relaxation. An important consequence being the finiteness of DC transport coefficients.
The presence of impurities is a generic instance where translational invariance is broken resulting in momentum dissipation.
Even more generically, a background lattice implies that, because of Umklapp scattering processes,
momentum is conserved only modulo reciprocal lattice vectors.

In general it is impossible to over-estimate the value of improved descriptions of impurities effects on the transport phenomena.
Indeed these are directly connected to the effects of disorder which are ubiquitous and important across the whole
condensed matter context. With this in mind, we mean to investigate a gauge/gravity model that describes 
a crossover from a weak-disorder quantum-critical regime to a disorder-dominated Fermi-liquid-like regime.
A crossover of this sort actually is expected in graphene \cite{sachdev}. At the outset, however, it is important to underline that the applicability
of a thermo-electric, momentum dissipating model in gauge/gravity is significantly wider. In fact, 
apart from impurity disorder, the presence of the lattice also breaks translational invariance.
As far as the present analysis is concerned, we focus on the effects of disorder neglecting the presence of a lattice.
This corresponds to interpreting the momentum dissipation as exclusively due to non-dynamical impurities (elastic scattering).
From the physical viewpoint, this possibly matches the expectations for graphene as long as the effects of phonons can be neglected
(this is actually the physical circumstance we are interested in)%
\footnote{To have a recent gauge/gravity instance where disorder is directly studied see \cite{Arean:2013mta}.}.

Describing momentum dissipating effects in the gauge/gravity framework is not an easy task. 
Actually all the early works applying the gauge/gravity correspondence to model condensed matter systems do not include momentum dissipation
and therefore feature a delta function at $\omega=0$ in the real part of the transport coefficients \cite{Hartnoll:2009sz}.
At present, several ways to introduce momentum relaxation in $AdS$/CFT are known. One possible approach is to consider spatially modulated 
backgrounds which directly simulate, for example, a lattice potential \cite{Horowitz:2012ky,Horowitz:2013jaa,Donos:2012js,Hartnoll:2008hs}.    
Another viable way consists in analyzing circumstances where a few light charged excitations scatter 
and dissipate their momentum on a bath of heavy neutral degrees of freedom \cite{Karch:2007pd,Hartnoll:2009ns,Faulkner:2010da}.

Recently it was proposed to introduce momentum relaxation in holography in an effective way 
(i.e. without a precise dynamical mechanism of momentum dissipation in mind) by using massive gravity \cite{Vegh:2013sk}. 
Indeed the bulk graviton mass breaks explicitly the diffeomorphism invariance of the gravitational action which 
in turn implies that the stress-energy tensor of the dual field theory is not conserved and momentum can be dissipated. 
This effective mechanism is theoretically extremely interesting since, in contrast to the methods featuring explicit 
spatial modulations, allows us to obtain quantitative information about correlators and physical observables 
without the need to resort to complicated numerical methods that typically involve the solution of systems of coupled 
partial differential equations%
\footnote{An analogous simplification occurs in \cite{Donos:2012js} for bulk dimensions $D=5$ and, more in general, in \cite{Donos:2013eha} for arbitrary $D$.}.
In \cite{Vegh:2013sk} the massive gravity model originally introduced in \cite{Hassan:2011vm} has been considered
within the holographic framework. In the bulk model the graviton mass is introduced by coupling the dynamical metric 
with a fixed fiducial metric that breaks diffeomorphism invariance. 
The way in which diffeomorphisms are broken depends on the particular choice of the fiducial metric.

In this paper (inspired by \cite{Vegh:2013sk} and subsequent articles) we analyze a massive gravity model where diffeomorphism 
invariance is broken in such a way that the dual field theory at the boundary conserves the energy but dissipates the momentum. 
In the condensed matter framework this kind of mechanism occurs, for example, in the presence of elastic electron 
scattering due to fixed impurities.
The same model has been studied also in \cite{Davison:2013jba,Blake:2013bqa}; specifically, in \cite{Davison:2013jba}, 
by analyzing the poles of the correlation functions in the hydrodynamic limit 
%(namely $T \gg \tau^{-1}$, where momentum is an almost conserved quantity),
%{\color{red}(namely at sufficiently high $T$, where momentum is an almost conserved quantity)},
(namely at sufficiently low momentum dissipation rate $\tau^{-1}$, where momentum is an almost conserved quantity), 
it was discussed that massive gravity is the dual gravitational realization 
of a system in which the conservation law for the stress-energy tensor is
\begin{equation}
\label{cons2}
\partial_t T^{tt}=0, \qquad  \partial_t T^{ti}= -\tau^{-1}T^{ti} \ ,
\end{equation}
where $\tau^{-1}$ is the momentum dissipation rate determined in terms of the graviton mass and the equilibrium thermodynamical quantities.
A precise specification of the validity range of the hydrodynamical treatment will be provided later.

In \cite{Blake:2013bqa} a universal analytical formula for the DC electrical conductivity 
in holographic massive gravity models was found. 
Comparing this expression with the electrical conductivity for a general hydrodynamic theory 
including the effects of impurity scattering (obtained in \cite{Hartnoll:2007ih}), 
it was noted that the two expressions agree provided that the scattering 
rate $\tau^{-1}$ assumes the specific form 
\begin{equation}
\label{taumassive1}
\tau^{-1}=-\frac{\mathcal{S} \beta}{2 \pi (\mathcal{E}+P)} \ ,
\end{equation}
found in \cite{Davison:2013jba}. We will be later more precise about 
the explicit expression of the scattering rate. For now it is sufficient 
to say that ${\cal S}$, ${\cal E}$ and $P$ are respectively the entropy
density the energy density and the pressure of the system at equilibrium and 
that $\beta$ is a parameter related to the bulk graviton mass which dually 
accounts for the ``strength'' of momentum dissipation.

Concerning the holographic massive gravity model at hand, some natural and important questions arise. 
The hydrodynamic theory considered in \cite{Hartnoll:2007ih} for a relativistic model with 
scattering due to impurities provides us universal expressions for the full set of thermo-electric transport coefficients in 
terms of the momentum dissipation rate $\tau^{-1}$ and of the thermodynamical quantities of the system. Is then the hydrodynamic 
regime of massive gravity completely consistent? In other words, given the expression for the scattering rate 
\eqref{taumassive1} and the thermodynamical quantities of the system, do all the transport coefficients agree with 
those obtained in \cite{Hartnoll:2007ih} from the hydrodynamical analysis?
Moreover, since massive gravity introduces the dissipation of momentum in an effective way, what is the physical character 
of the model when the hydrodynamical approximation ceases to hold? And, relatedly, is it possible in this non-hydrodynamic regime to understand anything 
concerning the possible microscopic processes giving rise to the effective mechanism of momentum relaxation?

As regards the microscopic realization of massive gravity, some progress has been attained in \cite{Kiritsis:2006hy,Aharony:2006hz,Apolo:2012gg,Andrade:2013gsa}, 
where it was investigated how massive gravity can be derived from general relativity in $AdS$, and in \cite{Blake:2013owa,Lucas:2014zea} where it was 
proven that a perturbative background lattice and random disorder provide a mass for the graviton. Nevertheless the ultimate answer about the 
microscopic origin of these massive gravity models is still hazy. We attempt to participate to such a debate 
from a rather phenomenological perspective. This primarily requires a full characterization of the behavior of the system
within the hydrodynamic regime, to make a consistency check, and outside hydrodynamics, to actually investigate its 
proper peculiarities. To this end, it is important to realize what can be understood about massive gravity and its holographic 
dual interpretation by analyzing the full set of thermo-electric transport coefficients. 

In order to compute the whole set of transport coefficients, we here rely completely on numerical methods. 
Actually, after the submission of the present paper, we obtained the DC thermo-electric transport coefficients also
analytically \cite{Amoretti:2014mma}. However, the numerical method illustrated in the following Sections is relevant 
by itself since it presents some interesting technical peculiarities and it is presently the only known method to compute the spectral 
behavior of the transport coefficients in massive gravity. 

The proper definition of the thermo-electric transport coefficients within massive gravity has to be considered carefully. 
We have addressed technical difficulties which arise in the holographic renormalization procedure concerning the need of finite 
boundary counter-terms in order to avoid unphysical features in the transport coefficients. Such a possibility is a crucial test that
massive gravity has to pass in order to be regarded as a sound holographic model. Note in fact that massive gravity models are 
obviously considered in a fully bottom-up and phenomenological spirit, at least as long as a consistent microscopic derivation 
of the bulk model is lacking.
For this reason a systematic check of the consistency of the dual phenomenological picture as a whole is always in order. 
To a similar purpose in \cite{Blake:2013bqa} a careful study of the equilibrium thermodynamical consistency of 
holographic models with massive gravitons has been considered. We here pursue further the investigation 
with an analogous attitude and we study the full thermo-electric linear response. 
On top of that, as the system at hand features a coupled thermo-electric dynamics, the extraction of the pure 
electrical and pure thermal response requires an attentive analysis of the intertwined linear response of the system%
\footnote{A similar dynamical circumstance has been addressed for instance in \cite{Kaminski:2009dh}.}.

At sufficiently high temperature $T$
%where the {\color{red}dissipation rate} $\tau^{-1}$ is the smallest energy scale of the system and the 
where the hydrodynamic limit is satisfied%
\footnote{We specify the precise definition of the range of validity of the hydrodynamic description 
providing later explicit formul\ae\ (see Equation \eqref{fc}) involving the physical variables of the system.} 
, we find that not only $\sigma_{DC}$ but the full set of transport coefficients agree with those
predicted by the hydrodynamic theory analyzed in \cite{Hartnoll:2007ih}. Specifically, they acquire the following form:
\begin{equation}
\begin{split}
\label{coeffintro}
\sigma_{DC}  =  \frac{1}{q^2}+&\frac{\rho^2}{\mathcal{E}+P}\tau \ ,  \qquad s_{DC}  =   -\frac{1}{q^2} \frac{\mu}{T}+\frac{\mathcal{S} \rho}{\mathcal{E}+P}\tau \ ,\\
&\bar{\kappa}_{DC}  =   \frac{1}{q^2}\frac{\mu^2}{T}+\frac{\mathcal{S}^2 T}{\mathcal{E}+P}\tau \ ,
\end{split}
\end{equation}
where $\sigma$, $s$ and $\overline{\kappa}$ are respectively the electric conductivity, the Seebeck coefficient
and the thermal conductivity (at zero electric field); in addition, $\rho$ is the charge density, $\mu$ is the chemical potential
and $q$ is a free parameter of the gravitational Lagrangian. 

In the hydrodynamical regime the dissipation rate $\tau^{-1}$ decreases with the temperature as $T^{-1}$.
However, in the low-$T$ region the dissipation rate increases and eventually the hydrodynamic picture and
expressions \eqref{coeffintro} cease to be valid. Remarkably, in this non-hydrodynamical regime, the 
transport coefficients that we obtained are in agreement with those computed in \cite{sachdev} using a Boltzmann approach 
for Dirac fermions with fermion-fermion interactions and a dilute density of charged impurities, namely:
\begin{equation}
\begin{split}
\label{coeffintrobal}
\sigma_{DC}  =  \frac{1}{q^2}+&\frac{\rho^2}{\mathcal{E}+P}\tau \ ,  \qquad s_{DC}  =   \frac{\mathcal{S} \rho}{\mathcal{E}+P}\tau \ ,\\
&\bar{\kappa}_{DC}  =  \frac{\mathcal{S}^2 T}{\mathcal{E}+P}\tau \ .
\end{split}
\end{equation}
The calculations of \cite{sachdev} are performed in the large-doping regime where $\mu \gg T$; 
we refer in particular to formul\ae\ (6.4) and (6.5) in \cite{sachdev}.

Furthermore, by analyzing the transport coefficients \eqref{coeffintrobal} and considering the specific expressions 
of the thermodynamical quantities of the holographic massive gravity model, we find that in this $\mu \gg T$ limit the system has some features in common
with the disorder-dominated Fermi-liquid regime. In fact the Wiedemann-Franz ratio is approximatively constant in temperature
even though its numerical value depends on $\tau$ and in general it is not that predicted by the Fermi-liquid.
More specifically, we have that the electric conductivity is temperature independent while the thermal conductivity $\bar{\kappa}_{DC}$ goes linearly
to 0 as $T \rightarrow 0$ and is proportional to the heat capacity%
\footnote{For related studies on the behavior of the thermo-electric transport coefficients in strongly correlated systems 
see \cite{Hartnoll:2014lpa}. As regards the gauge/gravity framework, a detailed discussion of the thermo-electric properties can be found 
in the study of coherent and incoherent metals in \cite{Gouteraux:2014hca,Donos:2014uba}. Interestingly, in \cite{Donos:2014uba},
an expression for the DC conductivity at finite temperature was obtained.}. 
This remarkable behavior and the agreement of our formul\ae\ with those for Dirac fermions obtained in \cite{sachdev} are hints 
of the fact that, at least in the large-doping regime, massive gravity could possibly admit a quasi-particle description, even though a proof of 
this statement requires further detailed studies \cite{amoretti}.

The paper is organized as follows. In Section \ref{sectionmassless} we review the standard 
analysis of the thermo-electric response of a holographic model without momentum dissipation and admitting
asymptotically $AdS_4$ Reissner-Nordstr\"{o}m solution. We systematically consider its holographic renormalization;
the expert reader can however jump directly to Section \ref{massivesection} where the momentum dissipating system is addressed.
There the massive gravity model of interest is defined and studied in depth. Again, particular attention is 
paid to the precise renormalization procedure and definition of the transport coefficients. In Section \ref{results}
we present a detailed account of the numerical results and describe the phenomenological picture which arises from them.
We comment on the presence and the physical significance of different regimes where the system admits either
a hydrodynamic or a ballistic-like description.
Particular attention is paid to the relation of our model to the physics 
of the crossover between a quantum-critical and a Fermi-liquid regime expected in dirty graphene.
 Eventually Section \ref{conclu} contains concluding remarks and
an outline of many interesting future prospects.

\section{Thermo-electric transport without momentum dissipation}
\label{sectionmassless}

In the present Section we review the thermo-electric transport in a simple system without momentum dissipation, 
namely we discuss the holographic dual of the well-known 4-dimensional Einstein-Hilbert-Maxwell model on a Reissner-Nordstr\"{o}m 
$AdS$ black hole.  This review is meant to recapitulate tidily the details of the holographic renormalization and the definition 
of transport coefficients in the standard momentum-conserving systems. We will then be able to highlight in later sections the differences 
one encounters in treating massive gravity.

\subsection{Bulk solution}

Consider the simplest 4-dimensional gravitational model admitting asymptotically $AdS$ charged black hole solutions,
namely an Einstein-Hilbert-Maxwell theory. This corresponds to the action
\begin{equation}
\label{actionNP}
S_{\text{RN}} = \int d^4x \sqrt{-g}\left[\frac{1}{2\kappa_4^2}\left(R-\frac{\Lambda}{L^2}\right)-\frac{1}{4 q^2}F_{\mu \nu}F^{\mu \nu} \right]+\frac{1}{2 \kappa_4^2} \int_{z=z_{UV}} d^3 x\ \sqrt{-g_b} \  2 K \ ,
\end{equation}
where we have already included the Gibbons-Hawking boundary term, which is expressed in terms of the induced metric $(g_{b})_{\mu \nu}$ 
and the extrinsic curvature $K$ on the surface at $z=z_{UV}$. Actually $z_{UV}$ represents a UV cutoff that will be 
sent to zero in the final step of the holographic renormalization procedure. As it is well known (see for example \cite{Wald:1984rg})
the Gibbons-Hawking term is necessary in order to have a well-defined bulk variational problem. In the action \eqref{actionNP} $\Lambda=-6$ 
is the dimensionless cosmological constant measured in units of the $AdS_4$ radius $L$;
$\kappa_4$ and $q$ are respectively the gravitational and Maxwell coupling constants and their dimension is
$[\kappa_4]=1$ and $[q]=0$.

From the action \eqref{actionNP} we get the Einstein and Maxwell equations
\begin{equation}\label{eom}
\begin{split}
&R_{\mu \nu}-\frac{g_{\mu \nu}}{2}\left(R-\frac{\Lambda}{L^2}\right)
=\gamma^2\left(F_{\mu \rho}F_{\nu}^{\; \rho}-\frac{g_{\mu \nu}}{4}F_{\rho \sigma}F^{\rho \sigma}\right) \ ,\\
&\partial_{\mu}\left(\sqrt{-g}F^{\mu \nu} \right)=0 \ ,
\end{split}
\end{equation}
where we have introduced the ratio of the gravitational and Maxwell couplings, namely $\gamma \equiv \frac{\kappa_4}{q}$.
Being the equations of motion \eqref{eom} insensitive to an overall rescaling 
of the action \eqref{actionNP}, they depend only on $\gamma$ and not on the individual couplings. 
It is worth noticing that for the simple model at hand $\gamma$ 
could be rescaled away by means of a field redefinition%
\footnote{Nevertheless this is not a general feature (e.g. it is not true for the holographic superconductor)
and we prefer to keep $\gamma$ explicit.}.

The model admits the following black-brane solution (see for example \cite{Tong}):
\begin{eqnarray}
\label{ansa}
&ds^2=\frac{L^2}{z^2} \left[-f(z) dt^2 + dx^2 + dy^2 + \frac{1}{f(z)} dz^2\right]\ , \qquad A=\phi(z)dt\ ,\\
\label{f_bg}
& f(z) = 1 - \left(1 + \frac{z_h^2 \gamma^2 \mu^2}{2 L^2}\right) \left(\frac{z}{z_h}\right)^3 
 + \frac{z_h^2 \gamma^2 \mu^2}{2 L^2} \left(\frac{z}{z_h}\right)^4 \ ,\\
\label{phi_bg}
& \phi(z) = \mu - q^2 \rho z
         = \mu \left(1 - \frac{z}{z_h}\right)\ .
\end{eqnarray}
where $z$ is the radial coordinate running from $z_{UV}$ at the UV radial shell to $z_h$ at the black hole horizon.
Of course, in the limit of vanishing cut-off, the radial UV shell is identified with the conformal boundary of 
the asymptotic $AdS$ geometry.

We recall that the coefficients of the leading and subleading near-boundary terms of the 
bulk gauge vector are respectively mapped to the dual chemical potential $\mu$ and charge density $\rho \equiv \mu/(q^2 z_h)$ of the 
corresponding global current in the boundary theory. 
Eventually, the black hole temperature (which coincides with that of the boundary theory) and the other thermodynamical quantities, 
such as the energy density $\mathcal{E}$ and the pressure $P$, can be derived in the standard holographic way 
(see for instance \cite{Hartnoll:2009sz,Musso:2014efa}). One obtains
\begin{equation}
\begin{split}
& T = -\left.\frac{1}{4 \pi} f'(z)\right|_{z=z_h}=-\frac{\gamma ^2 \mu ^2 z_h}{8 \pi  L^2} +\frac{3}{4 \pi z_h}\ ,\\
&\mathcal{E}=2P= \frac{ L^2}{z_h^3 \kappa^2_4}\left(1+\frac{z_h^2 \mu^2 \gamma^2}{2L^2}\right)\ .
 \end{split}
\end{equation}

\subsection{Fluctuations}

We consider vector fluctuations on the homogeneous and isotropic background specified by \eqref{ansa}, \eqref{f_bg} and \eqref{phi_bg}.
Without spoiling the generality of the treatment, the fluctuating fields that we study are the gauge field fluctuations along the $x$ spatial direction, namely $a_x$,
and the vector mode of the metric, $h_{tx}$; these are the relevant fluctuations in order to analyze the thermo-electric transport (see below). 
We further assume harmonic temporal dependence and isotropic spatial dependence (null momentum) for the fluctuations.

The fluctuation dynamics is governed by the Einstein and Maxwell equations \eqref{eom} which assume the following explicit form
\begin{eqnarray}
\label{a_eq}
 &a_x'' + \frac{f'}{f} a_x' + \frac{\omega^2}{f^2} a_x = - \frac{z^2\phi'}{f L^2}\left(h'_{tx} + \frac{2}{z} h_{tx}\right)\ ,\\
\label{h_eq}
 &h_{tx}' + \frac{2}{z} h_{tx} + 2 \gamma^2 \phi' a_x = 0\ ,\qquad \quad \qquad \; \; \; \;
\end{eqnarray}
where all the fields are functions of the $z$ variable alone and the primes denote derivatives with respect to $z$.
Despite the dynamics for the fluctuations $a_x$ and $h_{tx}$ is coupled, combining \eqref{a_eq}
and \eqref{h_eq} we obtain an equation where only $a_x$ and derivatives thereof appear,
\begin{equation}
\label{fluaxmassless}
 a_x''(z) + \frac{f'(z)}{f(z)} a_x'(z) + \left[\frac{\omega^2}{f(z)^2} -  2 \gamma^2 \  \frac{z^2\phi'(z)^2}{f(z)L^2}\right]  a_x(z) = 0\ .
\end{equation}
To actually solve the differential problem governing the fluctuation dynamics, we need
to specify appropriate boundary conditions at the horizon; we consider in-falling boundary conditions which are 
those needed to compute retarded correlators of the dual theory \cite{Kovtun:2005ev}. From \eqref{fluaxmassless}
we have that the gauge field fluctuations can be analyzed and solved without considering the metric fluctuations
which are later determined by means of \eqref{h_eq} upon substituting the solution for $a_x$. Therefore we have 
to impose the in-falling boundary conditions at the horizon on the gauge field alone,
\begin{equation}
a_{x}^{(\text{IR})}=(z_h-z)^{-\frac{i \omega}{4 \pi T}}(b_0+\mathcal{O}(z_h-z)) \ .
\end{equation}
Since the equation \eqref{fluaxmassless} is homogeneous, we can rescale the parameter $b_0$ to 1,
as a consequence $a_x$ and $h_{tx}$ are completely determined in terms of the frequency $\omega$ and the background
quantities. As we will see, this is not the case for massive gravity. There we face a system of two coupled equations 
where the ratio of the two leading IR coefficient of the fluctuation fields is physically relevant.
We will later discuss more in detail this important point.

\subsection{Renormalization of the fluctuation action}

In order to compute the correlators to be plugged into the Kubo formul\ae\ for the transport coefficients,
we need to consider the on-shell bulk action expanded at the second order in the fluctuations. 
The gauge/gravity prescription identifies the boundary value of the 
bulk fluctuation fields with the dual sources. The correlators of interest are then obtained 
taking appropriate functional derivatives of the on-shell action with respect to these sources.
This entire procedure represents the gauge/gravity version of the standard field theory 
paradigm to derive correlation functions.

In general the bulk on-shell action for the fluctuating field is divergent and needs to be properly renormalized. 
The holographic renormalization procedure consists in considering a regularized action to be integrated up to a near-boundary
radial cut-off; then, appropriate boundary counter-terms are considered and eventually the limit 
of zero cut-off defines the renormalized action.
The boundary counter-terms make the on-shell action finite once the UV cut-off goes to zero. They must respect the 
symmetries of the boundary theory and provide a well-defined bulk variational problem.
As mentioned before, in \eqref{actionNP} we have already added the Gibbons-Hawking boundary terms to the bulk action;
this provides a well-defined bulk variational problem for the fields. Then (see for instance \cite{Hartnoll:2009sz}) the only well-behaved 
boundary term needed in order to render the on-shell action finite is
\begin{equation}\label{ct}
 S_{\text{c.t.}} = \frac{1}{2 \kappa_4^2} \int_{z=z_{UV}} d^3 x\ \sqrt{-g_b}\ \frac{4}{L} \ .
\end{equation}
Eventually, the limit of vanishing cut-off is considered and (as we are interested 
in the linear response or, said otherwise, to two-point correlators) only the 
quadratic part of the action in the fluctuating field is retained. 
The renormalized quadratic action is defined as
\begin{equation}
S^{(2)}_{\text{ren}} = \left. \lim_{z_{UV} \rightarrow 0} S_{\text{RN}}+S_{\text{c.t.}}\right|_{\mathcal{O}(a_x, h_{tx})^2} \ .
\end{equation}

Once we have obtained a finite on-shell action, it is perfectly legitimate to ask ourselves
whether finite counter-terms could also be added. Such finite counterterms would lead
to ambiguities in the definition of the renormalized action%
\footnote{To have an example where finite counter-terms can be added to 
the bulk action of a holographic model and have an impact on the resulting physics, 
see \cite{Forcella:2014dwa}.}. We state once more that the counter-terms have to respect all the symmetries 
of the boundary theory%
\footnote{As a general feature, the correlators satisfy Ward identities related to the symmetries
of the model. In a generating functional framework, such identities (as the correlators themselves)
are obtained by appropriate functional derivatives of the generating functional itself and of the expectation values
of the various quantities in the theory. Counterterms (either finite or not) in the QFT action which respect 
the symmetries of the original theory affect both the Ward identities and the 
correlators in a consistent way \cite{Faddeev:1980be}.}, the power counting and the definition of the bulk variational problem.
This latter characteristic amounts to avoid introducing boundary terms containing radial derivatives. The former symmetry requirements impede us 
to consider terms as $a_i a^i$ which would brake the boundary gauge symmetry.
The power-counting criterion instead forbids us to consider $F_{ij}F^{ij}$
which is allowed by the symmetries but would force us to introduce new 
dimensionful parameters. We further notice that a Chern-Simons term is always trivial 
on our background solutions as a consequence of spatial rotational invariance. 
Such arguments exhaust all the possibilities
as far as the gauge field is concerned. Turning our attention to the metric, we are allowed
to consider two kinds of terms: a boundary cosmological constant and a term proportional 
to the boundary Ricci scalar.
The first actually appeared in \eqref{ct}; the latter is null as the manifold transverse
to the radial coordinate $z$ is flat Minkowski space-time upon which we are considering
homogeneous configurations in the space coordinates (i.e. null momentum).

From an asymptotic study of the equations of motion we have that the boundary expansions 
of the fields $a_x$ and $h_{tx}$ are
\begin{equation}
a_x(z) = a_x^{(0)} + a_x^{(1)} \frac{z}{L} + ... \ , \qquad h_{tx}(z) = \frac{L^2}{z^2}\; h_{tx}^{(0)} + h_{tx}^{(1)} \frac{z}{L} + ...\ ,
\end{equation}
and consequently the renormalized quadratic on-shell action for the model at hand is given by
\begin{multline}\label{S2}
 S^{(2)}_{\text{ren}} = \int d^3 x \Big[  \frac{1}{2q^2 L}\, a_x^{(0)}(-\omega) a_x^{(1)}(\omega)\\
 - \frac{1}{2 \kappa_4^2} \frac{3}{L}\, h_{tx}^{(0)}(-\omega)h_{tx}^{(1)}(\omega)
 -  \frac{\mathcal{E}}{4}\ h_{tx}^{(0)}(-\omega) h_{tx}^{(0)} (\omega) \Big]\ + \ \left(\omega \leftrightarrow -\omega \right) \ ,
\end{multline}
where we have Fourier transformed with respect to the time coordinate.

We anticipate that, as opposed to the model just analyzed in which there are no finite boundary counter-terms which can 
be added to the regularized action, in the massive gravity case, as we will see, the explicit breaking of diffeomorphism 
invariance allows us to add to the action non-trivial finite counter-terms. These may (and actually do) affect the physical quantities and,
in particular, the transport coefficients.

\subsection{Review and definition of the transport matrix}

The generic transport coefficient $C_{XY}$ is defined through the Kubo formula
\begin{equation}
 C_{XY} = i\, \omega\, \frac{\delta^2 S^{(2)}}{\delta X \delta Y} 
        = - \frac{i}{\omega} G_{XY}\ ,
\end{equation}
where $X,Y$ indicate the (here unspecified) physical sources (e.g. $E$ or $\nabla T$) while
the correlator $G$ is the Green function obtained through functional differentiation of the 
quadratic on-shell action $S^{(2)}$ with respect to the sources $a^{(0)}$ and $h^{(0)}$.
We are interested in computing the thermo-electric transport coefficients which relate 
at linear order the heat flow $\mathcal{h}Q_x\mathcal{i}$ and the electric current $\mathcal{h}J_x\mathcal{i}$ 
to the electric field $E_{x}$ and the gradient of the temperature $\nabla_{x}T$ in the following way:
\begin{equation}\label{traspo}
\begin{pmatrix} \mathcal{h}J_x\mathcal{i}\\ \mathcal{h}Q_x\mathcal{i} \end{pmatrix}=\begin{pmatrix}
\sigma & s T\\ s T & \bar{\kappa} T \end{pmatrix}\begin{pmatrix}
E_{x}\\ -\nabla_{x}T/T
\end{pmatrix}\ ,
\end{equation}
where $\sigma$ is the electric conductivity, $s$ is the Seebeck coefficient and $\bar{\kappa}$ is the 
thermal conductivity at vanishing electric field%
\footnote{From \eqref{traspo} we have that the thermal conductivity at vanishing electric current 
$\kappa$ is related to $\bar{\kappa}$ as follows: $\kappa=\bar{\kappa}-s \sigma^{-1} s T$.}.

The connection between the bulk field fluctuations and the fluctuations of the physical 
quantities (i.e. between $A$, $h$ and $E$, $\nabla T$) is the following \cite{Hartnoll:2009sz,Herzog:2009xv}:
\begin{equation}
\label{identification}
a_x^{(0)}=-\frac{i}{\omega}\left(E_x+\mu \frac{\nabla_x T}{T} \right) \ , \qquad h_{tx}^{(0)}=\frac{i}{\omega} \frac{\nabla_x T}{T} \ .
\end{equation}
In particular, as explained in \cite{Hartnoll:2009sz}, in order for this identification to be valid the theory 
must be invariant at least under temporal diffeomorphisms. Indeed to relate the fluctuation $h_{tx}^{(0)}$
to a thermal gradient one relies on a temporal diffeomorphism ``gauge'' transformation. 
This is related to the fact that in the framework of thermal quantum field theory, the imaginary 
period of the complexified time coordinate corresponds to the inverse temperature. 
The temporal diffeomorphism invariance is naturally satisfied in the standard formulation of general 
relativity but might be not true for massive gravity where diffeomorphism invariance is explicitly broken. 
However, as we will see in Section \ref{massivesection}, the massive gravity model which we consider is 
invariant under diffeomorphism in the $t-z$ directions and therefore the relations \eqref{identification} still hold.

From \eqref{identification} we have the following relations among the corresponding functional derivatives
\begin{eqnarray}\label{derivs}
 \frac{\delta}{\delta E_x} &=& - \frac{i}{\omega} \frac{\delta}{\delta a_x^{(0)}}\ , \\
 -T \frac{\delta}{\delta \nabla_x T} &=& - \frac{i}{\omega} \left[ \frac{\delta}{\delta h^{(0)}_{tx}} - \mu \frac{\delta}{\delta a_x^{(0)}} \right]\ ,
\end{eqnarray}
where the partial derivatives with respect to the sources $a_x^{(0)}$ and $h_{tx}^{(0)}$ are to 
be taken keeping to zero the source upon which one does not differentiate.
We underline that the sources $a_x^{(0)}$ and $h_{tx}^{(0)}$ are independent quantities.
Stated this, in order to compute the explicit expressions of the transport coefficients in terms of the background quantities and the 
near-boundary fluctuations, we start taking double functional derivatives of the 
on-shell renormalized and quadratic action \eqref{S2}. Namely,
\begin{equation}
 \frac{\delta^2 S}{(\delta E_x)^2} = \left(-\frac{i}{\omega}\right)^2 \left[ \frac{1}{q^2 L} \frac{\delta a_x^{(1)}}{\delta a_x^{(0)}} \right]\ ,
\end{equation}
\begin{equation}
 - T \frac{\delta^2 S}{\delta (\nabla_x T) \delta E_x} = 
 - T \frac{\delta^2 S}{\delta E_x \delta (\nabla_x T) } =  
 \left(-\frac{i}{\omega}\right)^2 \left[ 
 - \frac{3}{2 L \kappa_4^2} \frac{\delta h_{tx}^{(1)}}{\delta a_x^{(0)}} - \mu \frac{1}{q^2 L} \frac{\delta a_x^{(1)}}{\delta a_x^{(0)}}\right]\ ,
\end{equation}
and
\begin{equation}
 (-T)^2 \frac{\delta^2 S}{(\delta \nabla_x T)^2} = 
 \left(-\frac{i}{\omega}\right)^2 \left[ - \frac{3}{\kappa_4^2 L} \frac{\delta h_{tx}^{(1)}}{\delta h_{tx}^{(0)}} 
 - \mathcal{E}  - \frac{\mu}{q^2 L} \frac{\delta a_x^{(1)}}{\delta h_{tx}^{(0)}}
 + \frac{3 \mu}{\kappa_4^2 L} \frac{\delta h_{tx}^{(1)}}{\delta a_x^{(0)}} + \frac{\mu^2}{q^2L} \frac{\delta a_x^{(1)}}{\delta a_x^{(0)}}\right]\ .
\end{equation}
We observe that the equation for the fluctuations of the gauge field is independent of $h_{tx}$ and that, 
because of equation \eqref{h_eq}, $h_{tx}^{(1)}$ is completely determined in terms of $a_x^{(0)}$ and the parameters of the background,
\begin{equation}
h_{tx}^{(1)}=\frac{2}{3} \gamma^2 q^2 \rho L a_x^{(0)}\ ;
\end{equation}
hence we have%
\footnote{We anticipate that these relations are not valid in the massive gravity case where the fluctuations dynamics 
cannot be disentangled as in the massless gravity case.}
\begin{equation}
\label{relnull}
 \frac{\delta a_x^{(1)}}{\delta h_{tx}^{(0)}}=\frac{\delta h_{tx}^{(1)}}{\delta h_{tx}^{(0)}} = 0 \ \ \ \ \ \ \ \ \ \
 \text{and} \ \ \ \ \ \ \ \ \ \
 \frac{\delta h_{tx}^{(1)}}{\delta a_{x}^{(0)}}=\frac{2}{3} \gamma^2 q^2 \rho L \ .
\end{equation}
Eventually, we have that the entries of the transport matrix \eqref{traspo} are all 
expressible in terms of the background quantities and a unique electric conductivity \cite{Hartnoll:2009sz}:
\begin{equation}
 \sigma = - \frac{i}{\omega} \frac{1}{q^2 L} \frac{\delta a_x^{(1)}}{\delta a_x^{(0)}}\ , \ \ \ \ \ \ \
 s  = \frac{i}{\omega T} \rho - \frac{\mu}{T} \sigma\ ,
\end{equation}
and
\begin{equation}
\label{kappaRN}
 \bar{\kappa}  = - \frac{i}{\omega T} \left(
 - \mathcal{E}
 +  2 \mu \rho \right)
 + \frac{\mu^2}{T} \sigma\ .
\end{equation}

It is interesting to consider the thermal conductivity $\bar{\kappa} $
for a neutral black hole, namely for $\mu=0$. The only surviving contribution
is the imaginary pole whose residue is proportional to $\mathcal{E}$.
Relying on the Kramers-Kronig relations this corresponds to a delta function
at zero frequency in the real part of $\bar{\kappa} $ which 
encodes the lossless heat transport through a momentum conserving medium
induced by a thermal gradient.

\begin{figure}[H]
\centering
\includegraphics[width=7.3cm]{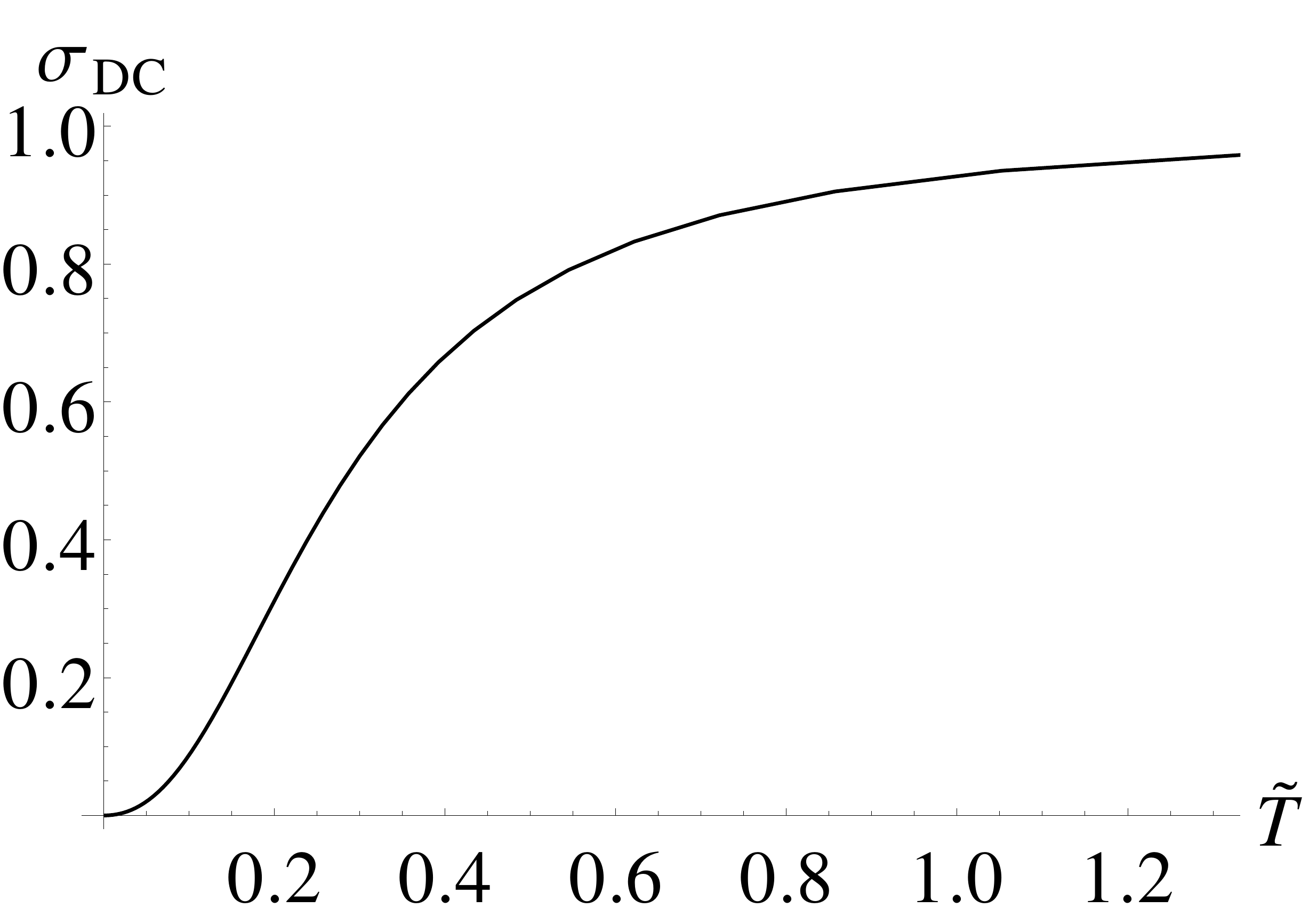}
\caption{The static limit of the electric conductivity $\sigma_{DC}$ as a function of the scale invariant temperature $\tilde{T}$. The values of the parameter of the model are: $\mu=1, \; L=1$ and $\gamma=1$.}
\label{sigmareissner}
\end{figure}

To conclude this brief review we plot in Figure \ref{sigmareissner} the static limit of the electric conductivity 
$\sigma_{DC}=\lim_{\omega \rightarrow 0} \sigma(\omega)$ as a function of the scale invariant temperature $\tilde{T}=T/\mu$.
Usually in the gauge/gravity literature one mostly discusses the spectral properties of the electric conductivity
without focusing on the temperature dependence. Instead, since we are concerned with the thermo-electric properties,
we find it interesting to study the temperature dependence of the static limits of all the transport coefficients,
i.e. also the Seebeck and the thermal conductivity. In fact, the experimental and real condensed matter investigations
of the thermo-electric and thermal coefficients are usually more focused on temperature dependence rather than the spectral behavior.
We have plotted the static electric conductivity for $\gamma=1$ and $\mu=1$, however its behavior for different values of the parameter 
$\gamma$ is the same since, as noted before, $\gamma$ can be reabsorbed through a field redefinition, namely the system is invariant under 
the scaling $\mu \rightarrow a \mu$ and $\gamma \rightarrow \gamma/a$.

It is important to note that, since there is a $\delta(\omega)$ in the real part of the conductivity, 
the static conductivity is defined as the limit for $\omega \rightarrow 0$ of the spectral conductivity
disregarding the delta function. From the numerical point of view, this coincides with computing the spectral 
conductivity at a value of $\omega$ much smaller than all the other scales in the system.
From Figure \ref{sigmareissner} we observe that the behavior of $\sigma_{DC}$ presents two regimes;
the ``cross-over'' region corresponds roughly with the energy scale set by the chemical potential $\mu$.
The two above-mentioned regimes consist in the following two behaviors: for $\tilde{T} \ll 1$ the static 
conductivity goes to zero quadratically while for $\tilde{T} \gg 1$ it saturates to $1/q^2$.

\section{Thermo-electric transport in massive gravity}
\label{massivesection}

In this Section, after discussing the basic properties of the massive gravity model which we consider, 
we will explain how to compute the thermo-electric transport coefficients for this system; the detailed 
analysis of the numerical results that we obtained is postponed to Section \ref{results}.
Unless specified otherwise, we refer to Section \ref{sectionmassless} for conventions and definitions. 

\subsection{The massive gravity model}

The idea underlying the application of massive gravity in holography consists in breaking the diffeomorphism 
invariance in the bulk by introducing a mass term for the graviton in such a way that one has momentum dissipation
in the boundary dual field theory.
Actually, several ways to give a mass to the graviton had been studied, but, following \cite{Vegh:2013sk}, 
we work here with the formulation of the massive gravity presented for the first time in \cite{Hassan:2011vm}.
The action of the model is: 
\begin{equation}
\label{massivelag}
\begin{split}
S = \int d^4x\ & \sqrt{-g} \left[ \frac{1}{2 \kappa_4^2}\left(R+\frac{6}{L^2}+\beta \left([\mathcal{K}]^2
-[\mathcal{K}^2]\right) \right)-\frac{1}{4 q^2}F_{\mu \nu}F^{\mu \nu} \right]\\
&+\frac{1}{2 \kappa_4^2} \int_{z=z_{UV}} d^3 x\ \sqrt{-g_b} \  2 K \ ,
\end{split}
\end{equation}
where $\beta$ is an arbitrary parameter having the dimension of a mass squared and the small square brackets denote a trace operation. 
Notice that the action \eqref{massivelag}
contains already the Gibbons-Hawking term necessary to have a well-defined bulk variational problem.
The matrix $(\mathcal{K}^2)^{\mu}_{\ \nu}$ is defined in terms of the dynamical metric $g_{\mu \nu}$ and a fiducial fixed metric 
$f_{\mu \nu}$ in the following way\footnote{ Within this formulation of massive gravity,
it is possible to consider also a linear term in the trace of $\mathcal{K}$; namely 
an $\alpha [ \mathcal{K}]$ term in the Lagrangian density where $\alpha$ is a numerical coefficient. 
However in this paper we always consider the case $\alpha = 0$. The reason for doing so is twofold: first a 
rigorous proof of the absence of ghosts in the model exists only in this $\alpha = 0$ case \cite{Vegh:2013sk}; 
secondly, as noted in \cite{Davison:2013jba}, with $\alpha \ne 0$ logarithmic terms appear in the near-boundary expansion of 
the bulk fields. The latter fact introduces non-standard divergences in the on-shell $2+1$-dimensional action.}
\begin{equation}
\begin{split}
&(\mathcal{K}^2)^\mu_{\ \nu}\equiv g^{\mu \rho}f_{\rho \nu} \ , \qquad \mathcal{K} \equiv \left( \sqrt{ \mathcal{K}^2} \right)^{\; \mu}_{\; \; \;\nu} .
\end{split}
\end{equation}
Along the lines of \cite{Vegh:2013sk}, we consider the following form for $f_{\mu \nu}$:
\begin{equation}
f_{\mu \nu}= \text{diag}(0,0,1,1) \ .
\end{equation}
Considering this particular form for the fiducial metric means that the action is still invariant under diffeomorphism 
in the $(z,t)$ plane, but not in the $(x,y)$ plane. At the dual level this implies that the theory has conserved energy 
but no conserved momentum.

At this point some comments are in order. In \cite{Davison:2013jba} it was proved that, in the limit of small momentum 
dissipation, namely when the temperature is greater than the characteristic momentum relaxation rate of the system, 
some observables computed in massive gravity are consistent with a hydrodynamical model which respects the modified conservation laws
given in \eqref{cons2}.
% \begin{equation}
% \label{cons1}
% \partial_t T^{tt}=0, \qquad  \partial_t T^{ti}= -\tau^{-1}T^{ti} \ ,
% \end{equation}
The $\tau$ appearing in the modified conservation relations is the characteristic momentum relaxation time of the system. 
Relations \eqref{cons2} coincide exactly with the conservation laws proposed in \cite{Hartnoll:2007ih} for a relativistic hydrodynamic 
model which includes impurity scattering in the limit of spatially isotropic perturbations.

% In order to understand better the microscopic mechanism which causes momentum relaxation in massive gravity 
% it is interesting to analyze the model also beyond the hydrodynamic approximation.
% In particular, one can argue to understand better the microscopic mechanism of massive gravity by analyzing the 
% full set of transport coefficients, which is exactly the aim of our analysis. 

\subsubsection{Background and thermodynamic}

The equations of motion descending from the action \eqref{massivelag} are:
\begin{equation}
\label{eqmotomasss}
\begin{split}
&R_{\mu \nu}-\frac{R}{2}g_{\mu \nu}+\frac{\Lambda}{2L^2}g_{\mu \nu}+X_{\mu \nu}=\gamma^2 T_{\mu \nu} \ ,\\
&\partial_{\mu}\left(\sqrt{-g}F^{\mu \nu} \right)=0 \ ,
\end{split}
\end{equation}
where $\gamma \equiv \frac{\kappa_4}{q}$ and
\begin{equation}
\begin{split}
&T_{\mu \nu}=F_{\mu \rho}F_{\nu}^{\; \rho}-\frac{g_{\mu \nu}}{4}F_{\rho \sigma}F^{\rho \sigma} \ ,\\
&X_{\mu \nu}= -\beta \left( \mathcal{K}^2_{\mu \nu}-2[\mathcal{K}]\mathcal{K}_{\mu \nu}+\frac{g_{\mu \nu}}{2}\left([\mathcal{K}]^2-[\mathcal{K}^2]\right) \right) \ .
\end{split}
\end{equation}
We want to study the system in the presence of a chemical potential, 
we then consider the same background ansatz as in \eqref{ansa}. In the massive case the black-brane solution is:
\begin{equation}\label{bbm}
\begin{split}
&\phi(z)= \mu - q^2 \rho z = \mu \left(1-\frac{z}{z_h} \right) \ , \qquad \rho \equiv \frac{\mu}{q^2 z_h} \ , \\
&f(z)=\frac{\gamma ^2 \mu ^2 z^4}{2 L^2 z_h^2}-\frac{\gamma
   ^2 \mu ^2 z^3}{2 L^2 z_h} -\frac{z^3}{z_h^3}-\frac{\beta 
   z^3}{z_h}+\beta  z^2+1 \ .
\end{split}
\end{equation}
In the limit $\beta \rightarrow 0$ the emblackening factor $f(z)$ reduces to that corresponding to the standard Reissner-Nordstr\"{o}m solution.
The black hole temperature is computed in the usual way leading to
\begin{equation}
\label{bhtm}
T=-\frac{f'(z_h)}{4 \pi}=-\frac{\gamma ^2 \mu ^2 z_h}{8 \pi  L^2} +\frac{\beta  z_h}{4 \pi }+\frac{3}{4 \pi z_h} \ .
\end{equation}
The full set of thermodynamical quantities were derived in \cite{Blake:2013bqa}. 
For the sake of later need, we report here the explicit expressions for the entropy density $\mathcal{S}$, 
the energy density $\mathcal{E}$ and the pressure $P$,
\begin{equation}
\label{thermomassive}
\mathcal{S}=\frac{2 \pi}{\kappa_4^2}\frac{L^2}{z_h^2} \ , \ \ \ \ \
\mathcal{E}=\frac{L^2}{z_h^3 \kappa_4^2} + \frac{L^2 \beta}{z_h \kappa_4^2} + \frac{\mu^2}{2 q^2z_h} \ , \ \ \ \ \
P=\frac{L^2}{2 \kappa_4 ^2 z_h^3}-\frac{\beta  L^2}{2 \kappa_4 ^2 z_h}+\frac{\mu ^2}{4 q^2 z_h} \ .
\end{equation}
Notice that the dual theory of a massive gravity has in general ${\cal E}\neq 2P$. The equation of state 
${\cal E} = 2P$ is expected for a $2+1$ dimensional conformal theory but, as it happens with the conservation laws
of the stress-energy tensor, the massive gravity set-up introduces modifications that are proportional to the 
mass parameter $\beta$.

\paragraph{Scales and scalings} \noindent $ \; $\\
As we have just noted observing the thermodynamic quantities, the massive parameter $\beta$ 
introduces a new scale in the model. This new scale affects the scaling symmetries of the bulk fields.
In fact, if we rescale the radial coordinate $z$ as $z \rightarrow a z$, 
we find that the other quantities of the model must scale as
\begin{equation}
(t,x,y) \rightarrow a (t,x,y) \ , \qquad \beta \rightarrow \frac{\beta}{a^2} \ , \qquad \mu \rightarrow \frac{\mu}{a} \ , \qquad z_h \rightarrow a z_h 
\end{equation}
in order for this scaling to be a symmetry of the action.
In particular, if we consider the scale invariant temperature  $\tilde{T} \equiv T/ \mu$ we find from \eqref{bhtm} that 
this is a function of the scale invariant quantities $\beta/ \mu^2$ and $ \mu z_h$:
\begin{equation}
\tilde{T} \equiv \frac{T}{\mu}= F \left( \frac{\beta}{\mu^2}, z_h \mu \right) \ .
\end{equation}
Moving the temperature while keeping fixed both the chemical potential $\mu$ and the 
mass parameter $\beta$ (which, as we will see, is related to the momentum dissipation rate in the dual field theory)
corresponds to varying the horizon radius $z_h$.

Finally, we note that, as in the massless case, the constant $\gamma$ can be rescaled away from the action \eqref{massivelag} 
by means of a redefinition of the gauge field. In fact the system is invariant under the scaling 
\begin{equation}
\label{scalingg}
\gamma \rightarrow a \gamma \ , \qquad \mu \rightarrow \mu / a \ ,
\end{equation}
namely the same scaling symmetry found in the Reissner-Nordstr\"{o}m $AdS$ black hole. This scaling affects in particular 
the transport coefficients and consequently to compute the transport coefficients at different values of $\gamma$ is equivalent 
to compute the same quantities at the corresponding rescaled values of the chemical potential.

\subsection{Fluctuations and transport in the massive case}

\subsubsection{Linearized equations and asymptotic expansions}

In order to obtain the transport coefficients, we need to expand the action \eqref{massivelag} 
at the second order in the fluctuation fields.
As in the massless bulk gravity case, we work in the zero momentum limit%
\footnote{For non-zero momentum the set of coupled fluctuations involves further components of the dynamical metric.
This has been studied in \cite{Davison:2013jba}.}. 
However, as opposed to the massless case, here the equations for $h_{tx}$ and $h_{zx}$ are independent 
and then we have to turn on both the fluctuations to be consistent. 
Hence we consider the following set of fluctuations
\begin{equation}
\label{flucmass}
\begin{split}
&A \rightarrow A+ e^{-i \omega t}\, a_x(z)\, dt \ ,\\
&ds^2 \rightarrow ds^2+2 e^{-i \omega t}\, h_{zx}(z)\, dz\, dx+2 e^{-i \omega t}\, h_{tx}(z)\, dt\, dx \ .
\end{split}
\end{equation}
Expanding the equations of motion \eqref{eqmotomasss} to the linear order in the fluctuations \eqref{flucmass} we obtain:
\begin{equation}
\label{eqeflu}
\begin{split}
&h'_{tx}+\frac{2}{z}h_{tx}+i \omega h_{zx} + 2 \gamma^2 \phi' a_x+ 2 \beta\, \frac{i f}{\omega}  h_{zx}=0 \ ,\\
&\frac{d}{dz}\left[h'_{tx}+i \omega h_{zx}+\frac{2}{z}h_{tx} + 2 \gamma^2 \phi' a_x \right]+ 2 \beta  \,\frac{h_{tx}}{f}=0 \ , \\
&\frac{d}{dz}\left(fa_x' \right)+\frac{\omega^2}{f}a_x + \frac{\phi' z^2}{ L^2}\left(h'_{tx}+\frac{2}{z}h_{tx}+i \omega h_{zx} \right)=0 \ .
\end{split}
\end{equation}
There are no derivatives of $h_{zx}$ in the first equation of motion which therefore can be algebraically solved
to obtain $h_{zx}$. We then substitute the solution inside the second equation. Finally we are left with two coupled equations for $a_x$ and $h_{tx}$:
\begin{equation}
\label{massflu1}
\begin{split}
&\frac{d}{dz} \left[f a_x' \right] + \frac{2\phi' z}{L^2}\ \frac{-\gamma^2 \phi' \omega ^2 z a_{x} + \beta f \left(z
   h_{tx}'+2 h_{tx}\right)}
   {2 \beta f  +\omega ^2 }
   +\frac{\omega ^2 }{f} a_{x} =0 \ , \\
&\frac{d}{dz} \left[\frac{ f}{z}\ \frac{2 \gamma ^2 \phi' z a_{x} +z h_{tx}'+2 h_{tx}}
   { 2 \beta f + \omega ^2 } \right]+\frac{ 1  }{f} h_{tx} =0 \ .
\end{split}
\end{equation}
In the $\beta \rightarrow 0$ limit the first equation in \eqref{massflu1} reduces to \eqref{fluaxmassless} 
obtained in standard massless gravity. This, however, cannot be simply interpreted as the fact that the fluctuation
dynamics in the limit $\beta \rightarrow 0$ coincides with that arising in the massless gravity on the Reissner-Nordstr\"{o}m black hole. 
Indeed, the second equation in \eqref{massflu1} 
shows that the limits $\beta \rightarrow0$ and $\omega \rightarrow 0$ do not commute. 
Since we are interested in computing DC observables we always consider the $\omega \rightarrow 0$ first.

\paragraph{IR expansion} \noindent \\
As usual, in order to compute the retarded correlators, we have to numerically solve the equations \eqref{massflu1} 
imposing the in-going wave boundary conditions at the horizon $z=z_h$, namely
\begin{equation}
\label{IWIC}
\begin{split}
&h_{tx}^{(\text{IR})}=(z_h-z)^{-\frac{i \omega}{4 \pi T}}(a_0+\mathcal{O}(z_h-z)),\\
&a_{x}^{(\text{IR})}=(z_h-z)^{-\frac{i \omega}{4 \pi T}}(b_0+\mathcal{O}(z_h-z)) .
\end{split}
\end{equation}
It is important to note that, unlike the case of fluctuations on pure Reissner-Nordstr\"{o}m black hole, it is impossible to combine the two equations 
\eqref{massflu1} in a unique equation for $a_x$. The dynamics of electric and thermal fluctuations is consequently more intimately mixed.
From the bulk standpoint, it is possible to rescale to 1 only one of the two coefficients $a_0$ and $b_0$. Said otherwise, the physics 
of the model is sensitive to the ratio $\eta= a_0/b_0$.
In the computations aimed at getting the transport coefficients, in order to isolate the purely electric response of the system, we have to tune 
the coefficient $\eta$ so that the thermal source vanishes. Symmetrically, to compute the pure thermal contribution, we must fix $\eta$ so that 
the electric field source is zero%
\footnote{In the context of mixed spin-electric transport a technically analogous situation arises in the unbalanced 
holographic superconductor \cite{Bigazzi:2011ak}.}.

\paragraph{UV expansion} \noindent \\
Near the boundary $z=0$ the expansion of the fluctuations in powers of $z$ is:
\begin{equation}
\label{espUV}
\begin{split}
&h_{tx}^{\text{UV}}(\omega, z)=\frac{L^2}{z^2}\left[h_{tx}^{(0)}(\omega)+\frac{1}{2}  (2 \beta + \omega^2)\frac{z^2}{L^2}h_{tx}^{(0)}(\omega)+\frac{z^3}{L^3}h_{tx}^{(1)}(\omega)+\mathcal{O}\left(\frac{z^4}{L^4}\right) \right]\ ,\\
&a_{x}^{\text{UV}}(\omega, z)=a_x^{(0)}(\omega)+\frac{z}{L}a_x^{(1)}(\omega)+\mathcal{O}\left(\frac{z^2}{L^2}\right)\ .
\end{split}
\end{equation}
The coefficients of the higher orders in the $z$ expansions can be determined in terms of the background parameters 
and the integration constants $h_{tx}^{(0)},h_{tx}^{(1)},a_x^{(0)}$, $a_x^{(1)}$. Since we are concerned with solutions 
of a system of second-order differential equations, these integration constants remain arbitrary in the UV analysis. 
As usual, once one imposes the above-mentioned IR boundary conditions at the horizon they are determined and can be read from the
full bulk solution. According to the standard holographic dictionary, we interpret $h_{tx}^{(0)}$ and $a_x^{(0)}$ as the sources of the
dual operators whose vacuum expectation values are given by $h_{tx}^{(1)}$ and $a_x^{(1)}$.

\subsubsection{On-shell action and renormalization}

The action \eqref{massivelag} diverges if evaluated on-shell at the quadratic order in the fluctuations. 
The counter-term which is necessary to make the quadratic action finite is, as in the massless case,
\begin{equation}
S_{\text{c.t.}}^{\text{(div)}} = \frac{1}{2 \kappa_4^2} \int_{z=z_{UV}} d^3 x\ \sqrt{-g_b}\ \frac{4}{L} \ .
\end{equation}
However, the reduced amount of symmetry in massive gravity allows one to introduce
additional finite counter-terms which are forbidden in the massless case.
More specifically, the larger freedom corresponds to the possibility of having terms that do not respect the spatial 
diffeomorphisms which are already broken by the graviton mass. Of course we still consider finite counter-terms 
which respect the power-counting (i.e. terms that do not require the introduction of further dimensionful coefficients), 
the (reduced) boundary symmetries and which lead to a well-defined bulk variational problem.

In accordance with the above-mentioned requirements, we are allowed to add only the following tower of finite counter-terms%
\footnote{We remind the reader that the case under consideration has zero spatial momentum $k$; hence
terms with spatial derivatives are automatically null. In such circumstances, terms involving the boundary Ricci scalar $R[\gamma]$
are vanishing as well.}
 \begin{equation}
 \label{fct}
 {\cal N} \int_{z=z_{UV}} d^3x\ z^{1-2n}\, \sqrt{-g_{b \; tt}}\, (g^{tt})^{n+1} g^{xx}\, (\partial_t^{(n)} h_{tx})(\partial_t^{(n)} h_{tx}) \ ,
\end{equation}
for all values of $n$. Here ${\cal N}$ is a normalization constant that depends on the dimensional parameters of the bulk theory%
\footnote{Not to be confuse with the $N\rightarrow\infty$ rank of the boundary theory gauge group.}. 
It is important to notice that for $n\neq0$ the counter-terms \eqref{fct} introduce polynomial contributions to the imaginary part 
of the $\mathcal{h}T_{tx}T_{tx}\mathcal{i}$ correlator and that such contributions diverge at large frequency. 
We exclude this behavior on the basis of field theoretical arguments on the high-$\omega$ behavior of physical correlators
and therefore we retain only the $n=0$ case, namely
\begin{equation}
\label{fctt}
 S_{\text{c.t.}}^{\text{(fin)}}(a) = \frac{a}{2}\ \mathcal{E} \int_{z=z_{UV}} d^3x\ \frac{z}{L}\, \sqrt{-g_{b \; tt}}\, g^{tt}g^{xx}\, h_{tx}h_{tx}\ ,
\end{equation}
where $a$ is a dimensionless parameter on which the finite counter-term depends. The freedom associated 
to the choice of a specific value for $a$ appears as a renormalization ambiguity of the model or, 
said otherwise, to a renormalization scheme dependence. However, in order to eliminate an unphysical delta function at $\omega=0$ 
in the thermal conductivity, we must choose $a=-\frac{1}{2}$. We will comment further on this important
aspect in the following Sections; here we anticipate the remark to underline that the physical model at hand 
is eventually not affected by renormalization ambiguities.
Terms similar to \eqref{fct} but containing $h_{zx}$ do not respect spatial translation invariance%
\footnote{One can recover spatial translations considering a spatial diffeomorphism 
where the coordinate variation $\xi$ is a constant. The component $h_{zx}$ has a non-vanishing 
variation contributed by the non-trivial Christoffel symbols involving the coordinate $z$.
Of course, interpreting the massive gravity model as an effective way to account for spatial inhomogeneities,
one would drop the spatial translation invariance requirement. In such circumstances it is possible that
wider classes of counter-terms could be considered. This analysis is however beyond the purpose of the present 
treatment.}.

The total on-shell action reduces to a purely boundary term. Fourier transforming the fields and substituting $h_{zx}$ 
by means of the second equation in \eqref{eqeflu} we obtain
\begin{equation}
\label{boundaryac}
 \begin{split}
S_{\text{tot}} &=S + S_{\text{c.t.}}^{\text{(div)}} + S_{\text{c.t.}}^{\text{(fin)}}\\
&=\lim_{z_{UV} \rightarrow 0}\ V \int \frac{d \omega}{2 \pi} 
\Big[ \frac{\mu  z^2 \left(\beta  f+\omega ^2\right)}{L^2 q^2 z_h \left(2 \beta  f+\omega ^2\right)}a_x h_{tx}+\frac{\beta  z^2 f }{2 \kappa_4 ^2 L^2 \left(2 \beta  f+\omega ^2\right)}h_{tx} h_{tx}'\\
&-\frac{f }{2 q^2}a_x a_x'
+\left(\frac{z }{2 \kappa_4 ^2 L^2 \sqrt{f}} +\frac{a\mathcal{E}z}{4L} \, \sqrt{-g_{b \;tt}}\, g^{tt}g^{xx}\,\right) h_{tx}h_{tx}
\Big]_{z=z_{UV}} \ + \left(\omega \leftrightarrow -\omega \right) \  ,
 \end{split}
\end{equation}
where the prime denote the derivative with respect to the radial variable $z$, the arguments of the first and second fluctuation 
in each pair are respectively $(-\omega,z)$ and $(\omega,z)$ and $V$ represents the volume of the spatial manifold.

The boundary action \eqref{boundaryac} evaluated on the boundary expansions \eqref{espUV} allows us to compute the transport coefficients, 
(for details on the computation of the transport coefficients see Appendix \ref{compumass}).

\subsubsection{Definition of the transport coefficients}

The computation of the transport coefficients is analogous to that illustrated for the massless case, but with two important differences.
The first one is that, since we are dealing with two coupled differential equations, relations \eqref{relnull} are not 
valid and we have to keep into account that:
\begin{equation}
 \frac{\delta a_x^{(1)}}{\delta h_{tx}^{(0)}} \ne 0 \ , \ \ \ \ \ \ \ \text{and} \ \ \ \ \ \ \
 \frac{\delta h_{tx}^{(1)}}{\delta h_{tx}^{(0)}} \ne 0\ .
\end{equation}
The second is that, on the computational level, in the massive case the IR parameter $\eta=a_0/b_0$ coming from the boundary conditions 
at the horizon \eqref{IWIC} has a physical relevance and cannot be simply rescaled to 1. Indeed we have to tune $\eta$ depending on which source we 
need to set to zero in performing the functional derivatives. We resort to a numerical shooting method to the purpose of 
finding the value of $\eta$ corresponding to the desired UV source set-up.

Finally, the explicit expressions of the electric conductivity $\sigma$, the thermal conductivity $\bar{\kappa}$ and the thermo-electric 
conductivity $s$ (obtained in Appendix \ref{compumass}) are
\begin{equation}
\sigma=-\left. \frac{1}{q^2L}\frac{i}{\omega} \frac{ \delta a_{x}^{(1)}}{ \delta a_x^{(0)}}\right|_{h_{tx}^{(0)}=0} \ ,
\end{equation}
\begin{multline}
\label{kappa}
\bar{\kappa} = -2\frac{i}{T \omega} \Big[(a-1)\frac{\mathcal{E}}{2} - \frac{3 \beta}{2\kappa_4^2 L(2\beta + \omega^2)} \left. \frac{ \delta h_{tx}^{(1)}}{ \delta h_{tx}^{(0)}} \right|_{a_x^{(0)}=0}
  -  \frac{\mu}{2q^2L} \left. \frac{ \delta a_{x}^{(1)}}{ \delta h_{tx}^{(0)}} \right|_{a_x^{(0)}=0} +\\  \frac{\mu^2}{z_h q^2} \frac{\beta + \omega^2}{2 \beta + \omega^2}
  +  \frac{3 \mu \beta}{2\kappa_4^2 L(2\beta + \omega^2)} \left. \frac{ \delta h_{tx}^{(1)}}{ \delta a_x^{(0)}} \right|_{h_{tx}^{(0)}=0}
  +  \frac{\mu^2}{2q^2L} \left. \frac{ \delta a_{x}^{(1)}}{ \delta a_x^{(0)}} \right|_{h_{tx}^{(0)}=0}  \Big]\ ,
\end{multline}
\begin{multline}
\label{seebeck}
s =-\frac{i}{T \omega}\Big[ \frac{1}{2q^2L} \left.\frac{\delta a_{x}^{(1)}}{\delta h_{tx}^{(0)}}\right|_{a_x^{(0)}=0}
 - \frac{\mu}{z_h q^2} \frac{\beta + \omega^2}{2 \beta + \omega^2}\\ - \frac{3 \beta}{2\kappa_4^2 L(2\beta + \omega^2)} \left.\frac{ \delta h_{tx}^{(1)}}{\delta a_x^{(0)}}\right|_{h_{tx}^{(0)}=0} 
 -   \frac{\mu}{q^2L} \left.\frac{\delta a_{x}^{(1)}}{\delta a_x^{(0)}}\right|_{h_{tx}^{(0)}=0} \Big] \ .
\end{multline}
As anticipated, the thermal conductivity $\bar{\kappa}$ depends explicitly on the parameter $a$ introduced by
the finite counter-term \eqref{fct} and, as explained in the next paragraph, we fix the value of $a$ according to physical requirements.

\paragraph{Fixing the finite counter-term} \noindent\\
As it is evident from \eqref{kappa}, only the imaginary part of the the thermal conductivity depends on the value of the parameter $a$. 
This parameter, which corresponds to the normalization of the finite counter-term \eqref{fct}, has a key role
in allowing us to get a sensible physical picture. For instance let us note that if we just set $a=0$ we find as the result 
of the numerical computations that the imaginary part of the thermal conductivity has a pole at $\omega=0$. 
The Kramers-Kronig relations map such a pole to a delta function $\delta(\omega)$ in the real part of $\bar{\kappa}$.
A delta function in the thermal conductivity describes a perfectly efficient (lossless) transport of heat which 
is unphysical given that we are concerned with a system that dissipates momentum.

The apparent inconsistency can be completely fixed by setting $a=-1/2$. Observe that the divergence in the imaginary part of 
$\bar{\kappa}$ is evidently contributed by the first term in \eqref{kappa} which diverges as $i(1-a) \mathcal{E}/ \omega T$.
Actually also the the second term in \eqref{kappa} yields an analogous contribution which, however, needs to be uncovered and treated
numerically. Indeed an attentive numerical analysis shows that
\begin{equation}
\label{divergence}
 \text{Im} \left( \frac{3 \beta}{\kappa_4^2 L(2\beta + \omega^2)T \omega} \left. \frac{ h_{tx}^{(1)}}{ h_{tx}^{(0)}} \right|_{a_x^{(0)}=0} \right) \sim -\frac{3}{2}\frac{ \mathcal{E}}{\omega T} \ .
\end{equation}
where the numerical factor in front of $\mathcal{E}$ does not depend (according to our numerical precision) 
on the particular value of the other parameters of the model.

Notice that the numerical result \eqref{divergence} seems to provide an analytical insight.
This noticeable conclusion is not only based on an accurate numerical treatment but on a theoretical expectation
as well. Recall that in the massless gravity set-up the lossless thermal transport of a neutral black hole is proportional
to the energy density ${\cal E}$. This feature can be regarded as a generic characteristic independent of the 
details of the particular holographic model one considers. Also in massive gravity, where the lossless thermal transport
would lead to unphysical consequences, we can reliably expect that it can be reabsorbed by means of tuning the coefficient
with which ${\cal E}$ appears in the thermal conductivity. This argument supports us in distilling an analytical conclusion 
from the numerical data.

\begin{figure}[h!]
\centering
\includegraphics[width=7cm]{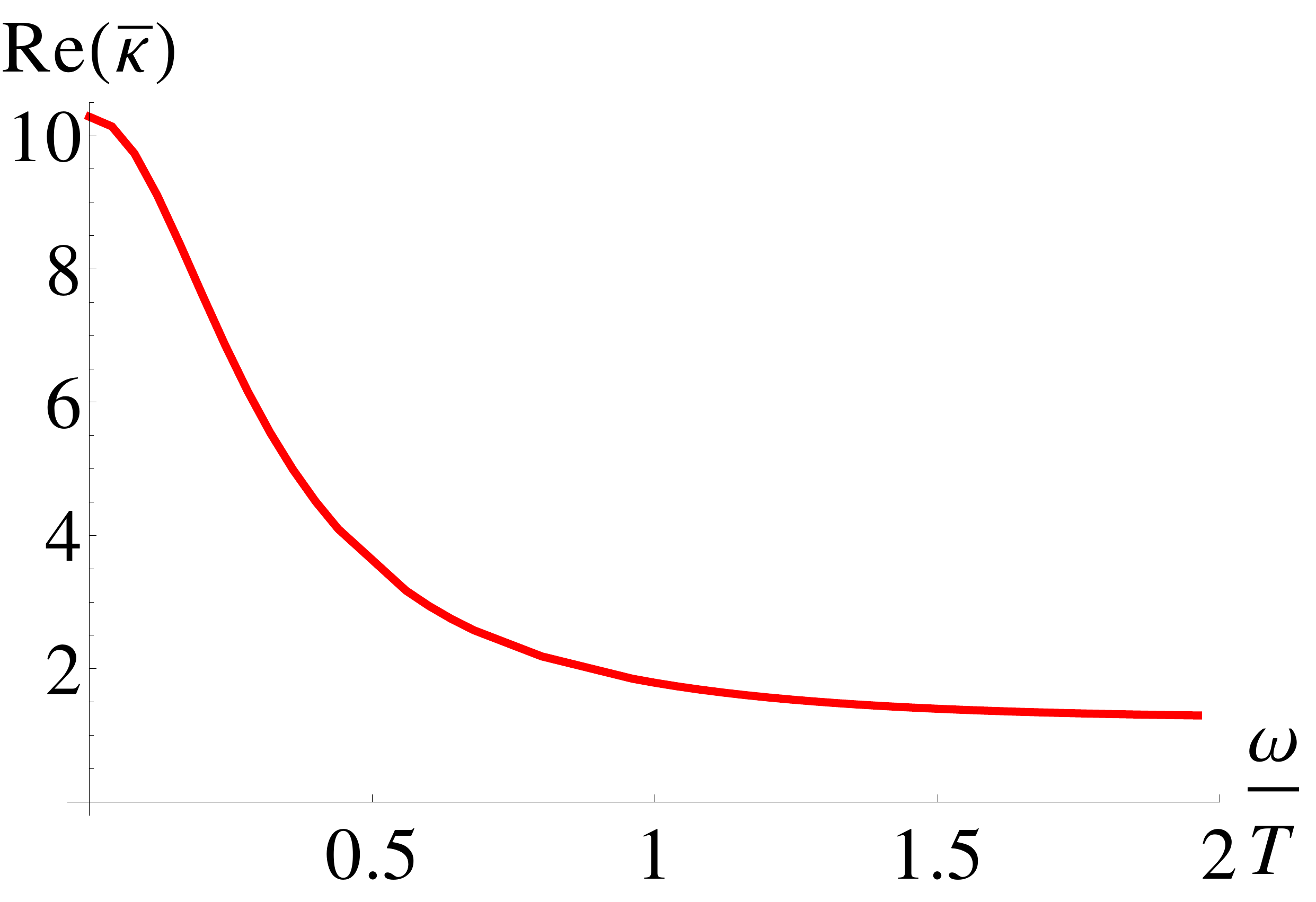}
\includegraphics[width=7cm]{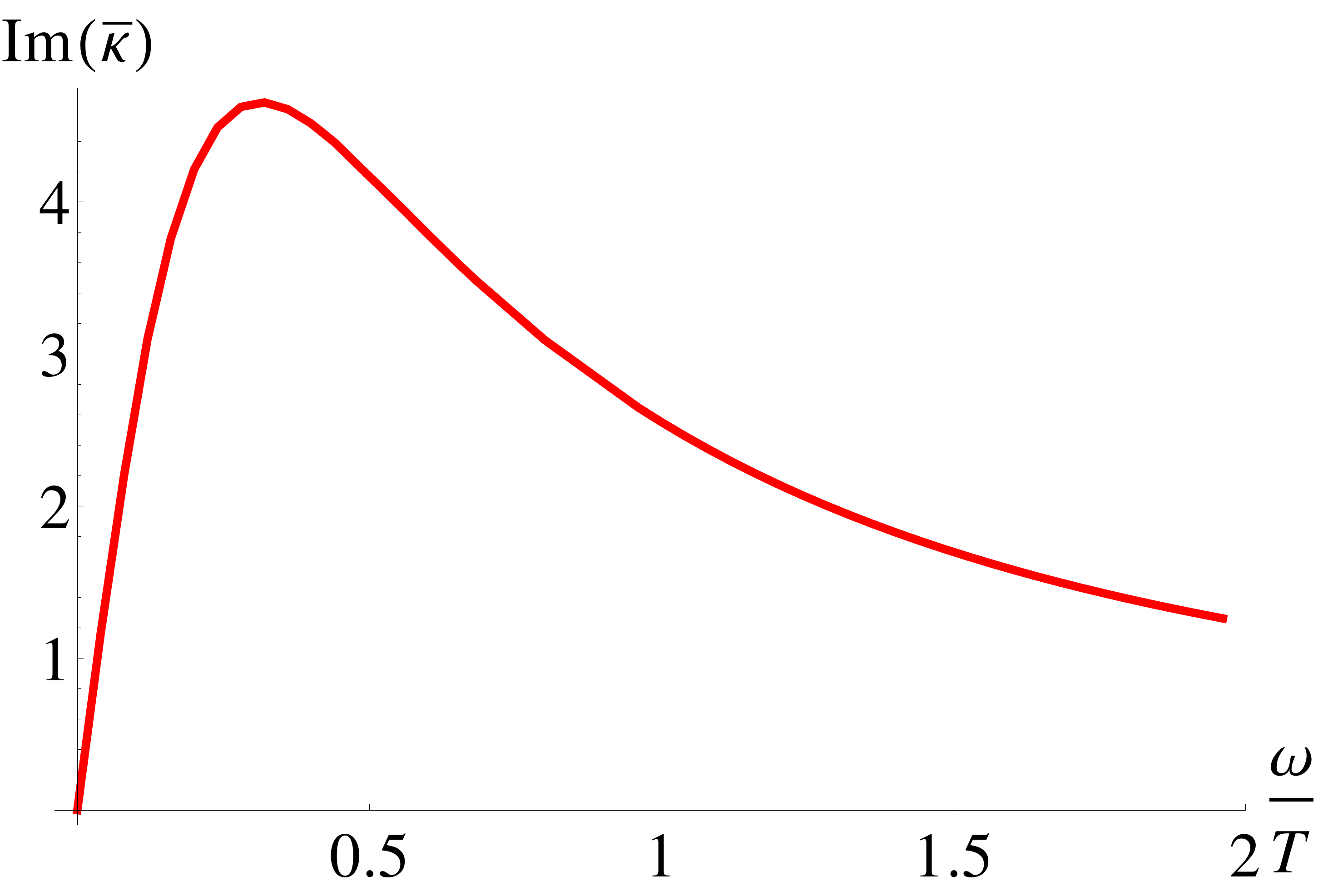}
\caption{Real (left) and imaginary (right) part of the thermal conductivity $\bar{\kappa}(\omega)$ for $\beta=-0.44$, $\gamma=0.6$ and $T/\mu=1$.}
\label{kappaspettr}
\end{figure}

Let us rely on the same point looking the details of the 
formul\ae. For small $\omega$ the imaginary part of the thermal conductivity behaves as
\begin{equation}
\text{Im} (\bar{\kappa}) \sim -\left(a+\frac{1}{2}\right) \frac{2\, \mathcal{E}}{\omega T} \ .
\end{equation}
If we set $a=0$ we find the same divergence as for massless gravity on the neutral black hole solution (see \eqref{kappaRN} with $\rho=0$). 
However, in the massive case the symmetries of the model allow us to consider $a \ne 0$, and in particular if we set $a=-\frac{1}{2}$ 
we find that the imaginary part of the thermal conductivity goes to zero as $\omega \rightarrow 0$.
as expected for the imaginary part of a physical transport coefficient in the presence of momentum dissipation and in the DC limit (see Figure \ref{kappaspettr}).

\section{Transport coefficients analysis}
\label{results}

\subsection{The dissipation rate and the hydrodynamic regime}

A general hydrodynamic theory in the vicinity of a quantum critical point where Lorentz invariance 
is weakly broken by the presence of weak disorder (associated to an impurity scattering rate $\tau^{-1}$)
was developed in \cite{Hartnoll:2007ih}.
Being the considered impurity scattering an elastic mechanism, \cite{Hartnoll:2007ih} supposes that weak disorder affects
only the momentum conservation while preserving the energy as encoded in \eqref{cons2}.
As a general result of this hydrodynamic approach, once the scattering rate $\tau^{-1}$ and the thermodynamical quantities% 
\footnote{See \eqref{thermomassive} for the explicit thermodynamical expressions of the thermodynamical quantities 
in the system at hand.} of the system
are provided, all the transport coefficients take the following form:
\begin{eqnarray}
\label{sigmaart}
\sigma_{DC} & = & \lim_{\omega \rightarrow 0} \sigma(\omega)=\sigma_Q+\frac{\rho^2}{\mathcal{E}+P}\tau \ , \\
\label{sart}
s_{DC} & = & \lim_{\omega \rightarrow 0} s(\omega)= -\sigma_Q \frac{\mu}{T}+\frac{\mathcal{S} \rho}{\mathcal{E}+P}\tau \ ,\\
\label{kappaart}
\bar{\kappa}_{DC} & = & \lim_{\omega \rightarrow 0} \bar{\kappa}(\omega)= \sigma_Q \frac{\mu^2}{T}+\frac{\mathcal{S}^2 T}{\mathcal{E}+P}\tau \ ,
\end{eqnarray}
where $\sigma_Q$ has to be determined in terms of a constitutive description of the system.

In \cite{Davison:2013jba, Blake:2013bqa} it was proven that massive gravity has a hydrodynamic regime 
when the scattering rate $\tau^{-1}$ is sufficiently small (holding all the other variables, e.g. the temperature and the
chemical potential, fixed; see \eqref{fc} in the following) and therefore the momentum conservation violation is small as well.
This regime is captured by the general hydrodynamic treatment described in \cite{Hartnoll:2007ih}.
In particular in \cite{Davison:2013jba}, by analyzing the poles of the correlators in such a hydrodynamic limit
and determining the scattering rate as
\begin{equation}
\label{taumassive}
\tau^{-1}=-\frac{\mathcal{S} \beta}{2 \pi (\mathcal{E}+P)} \ ,
\end{equation}
it was demonstrated that massive gravity is well described by the modified conservation law \eqref{cons2}. 
More precisely, \cite{Blake:2013bqa} provides an analytical expression for the static electric conductivity $\sigma_{DC}$ 
for every value of the temperature $T$ in the massive gravity model at hand,
\begin{equation}
\label{sigmamass}
\sigma_{DC}=\frac{1}{q^2}-\frac{\kappa_4^2 \rho^2}{L^2}\frac{z_h^2}{\beta} \ ,
\end{equation}
which agrees with \eqref{sigmaart} when the scattering rate $\tau^{-1}$ is given by \eqref{taumassive} and $\sigma_Q=1/q^2$.

% It is important to underline again that formul\ae\ \eqref{sigmaart}-\eqref{kappaart} and the argument followed in \cite{Davison:2013jba} 
% are expected to be valid only when the scattering rate is the smallest energy scale in the system, namely
% $\tau^{-1} \ll T$. The reason being that the modified conservation law \eqref{cons2}, 
% which is the starting point of \cite{Davison:2013jba} and \cite{Hartnoll:2007ih}, is accurate only if the dissipation rate 
% $\tau^{-1}$ and the frequency of the fluctuations $\omega$ are small compared to the characteristic scales of the system \cite{sachdev}.

\begin{figure}[h!]
\centering
\includegraphics[width=9cm]{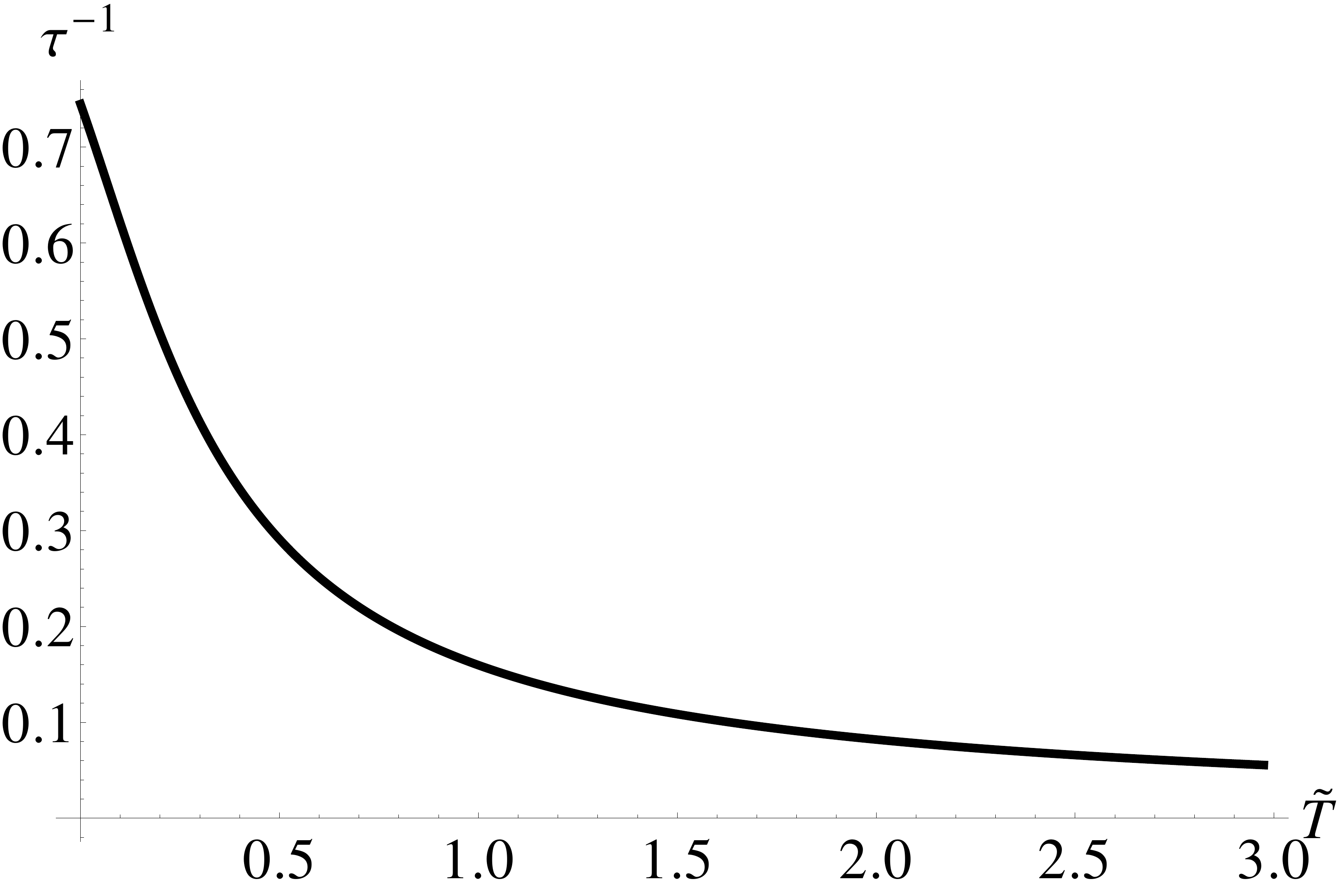}
\caption{Scattering rate $\tau^{-1}$ as a function of the scale invariant temperature $\tilde{T}$ for $\beta=-0.44$ , $\mu=1$, $\gamma=1$.}
\label{taumassfig}
\end{figure}

To have a complete picture of the behavior of the system at hand also beyond its hydrodynamical regime
we need to study the whole range of the scale invariant temperature $\tilde{T}=T/\mu$%
\footnote{As explained already in Section \ref{massivesection}, we recall that the correct way to vary the temperature corresponds
to move the horizon radius $z_h$ keeping fixed the chemical potential $\mu$. This because there are more independent scales in the system
in addition to $T$ and $\mu$, such as, for instance, $\beta$. As a consequence, to obtain the scattering rate as a function of the temperature, we have substituted 
$z_h(\tilde{T})$ in \eqref{taumassive}.}.
Keeping into account the expressions for the thermodynamical quantities given in \eqref{thermomassive} and 
writing the scattering rate \eqref{taumassive} as a function of $\tilde{T}$, we obtain the following limiting behaviors
for $\tilde{T}\ll1$ and $\tilde{T}\gg1$ 
\begin{equation}
\label{taumassfor}
\begin{split}
& \tau^{-1} \underset{\tilde{T} \ll 1}{=} -\frac{\beta  \sqrt{L^2  \left(\gamma ^2 \mu ^2-2 \beta  L^2 \right)}}{\sqrt{6} \gamma ^2 \mu ^2}-\frac{ \beta  \left(2 \pi 
   \beta  L^4 +\pi  \gamma ^2 \mu ^2 L^2 \right)}{3 \gamma ^4 \mu ^3}\tilde{T}+\mathcal{O}\left(\tilde{T}^2\right)\\
&\tau^{-1} \underset{\tilde{T} \gg 1}{=} -\frac{\beta  \left(2 \beta  L^2 -\gamma ^2 \mu ^2\right)}{2  \pi  \mu  \left(3 \gamma ^2 \mu ^2+2 \beta  L^2
   q^2\right)} \frac{1}{\tilde{T}}+\mathcal{O}\left(\tilde{T}^{-2}\right)
\end{split}
\end{equation}
We report them in Figure \ref{taumassfig}.
According to \eqref{taumassive} and Figure \ref{taumassfig}, the scattering time approaches a constant at $\tilde{T} \rightarrow 0$ 
and decreases as $\tilde{T}^{-1}$ when $ \tilde{T} \rightarrow \infty$. This qualitative behavior is the same for every allowed values
of the parameters of the model%
\footnote{We remind the reader that, as explained in \cite{Vegh:2013sk, Davison:2013jba},
$\beta$ must be negative in order for the scattering rate to be positive.}.
In the low $\tilde{T}$ region where $\tau^{-1}$ increases the hydrodynamic approximation is 
no longer accurate as we will explicitly see.

% Hence, there is always a region at low $\tilde{T}$ where $\tau^{-1}$ is greater than the temperature and in such region 
% the hydrodynamic approximation is no longer accurate. 

As already noted, the conductivity \eqref{sigmamass} obtained in \cite{Blake:2013bqa} is valid for every value of the temperature $T$. 
The passages performed in \cite{Blake:2013bqa} leading from the expression of the conductivity \eqref{sigmamass} to that
of the scattering rate \eqref{taumassive} are performed within the hydrodynamical regime. Our purpose is at first 
to cross-check the validity of the hydrodynamical approximation by comparing our numerical results regarding the other transport coefficients 
with \eqref{sart} and \eqref{kappaart}. Then, adopting the same expression for the scattering rate \eqref{taumassive}
also beyond the hydrodynamical region, we are interested in characterizing the behavior of the system in the whole 
temperature range (see Subsection \ref{beyond}). 

For the moment being we stick to the hydrodynamical regime. As regards the electric conductivity,
keeping into account that for the holographic model at hand the charge density is $\rho=\frac{\mu}{q^2 z_h}$, it is evident that $\sigma_{DC}$ 
(see Equation \eqref{sigmamass})
does not depend on the horizon radius $z_h$ and then on the temperature%
\footnote{This represents a peculiarity of the model at hand. According to \cite{Blake:2013bqa}, the inclusion of a term $\alpha [ \mathcal{K} ] $
in the gravitational action \eqref{massivelag} would lead to a conductivity $\sigma_{DC}$ which actually depends on the temperature $T$.}.  
We have verified the correctness of our numerical computations by comparing the results for $\sigma_{DC}$ against the analytic formula \eqref{sigmamass}.

\begin{figure}[h!]
\centering
\includegraphics[width=7.3cm]{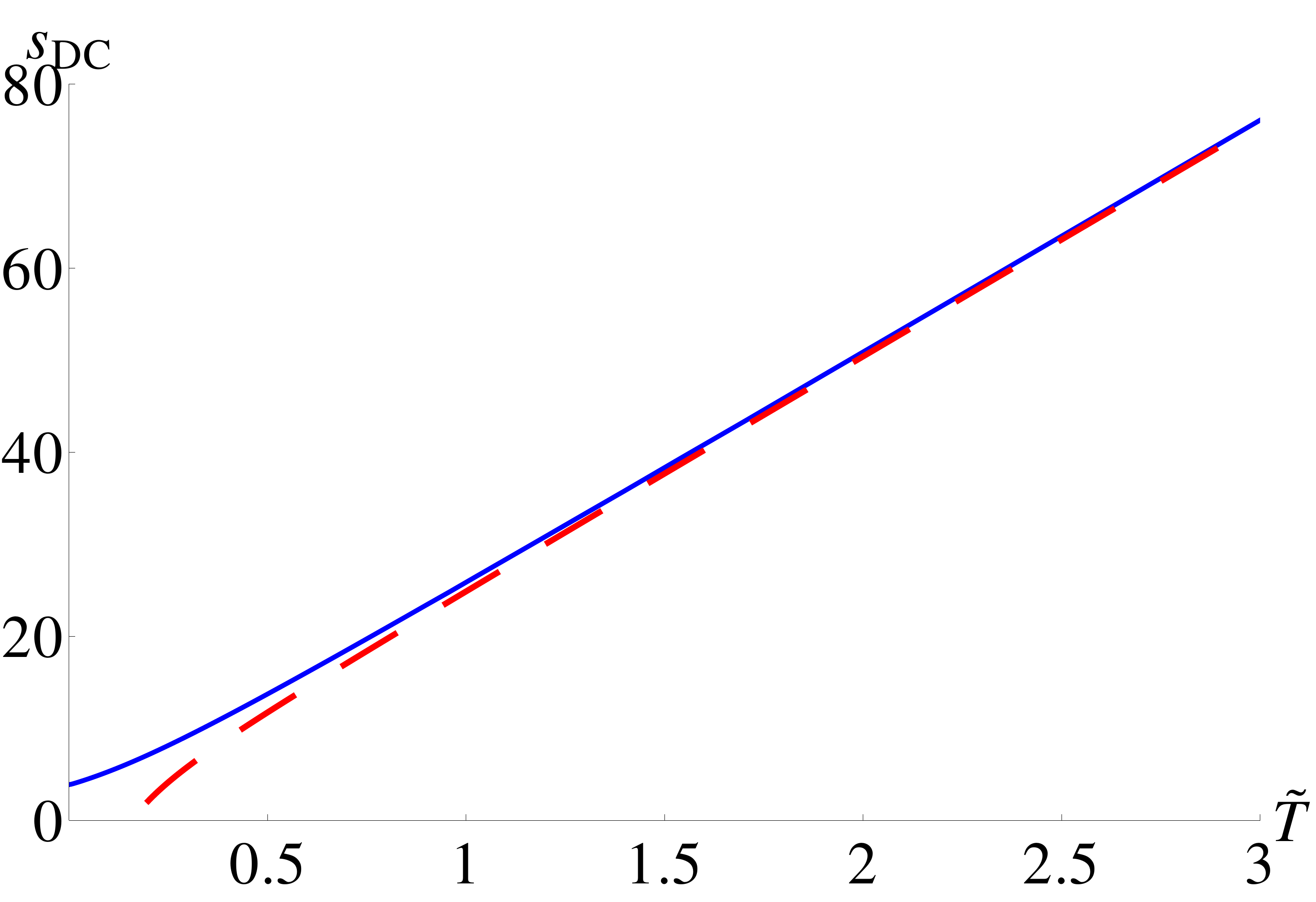}
\includegraphics[width=8.5cm]{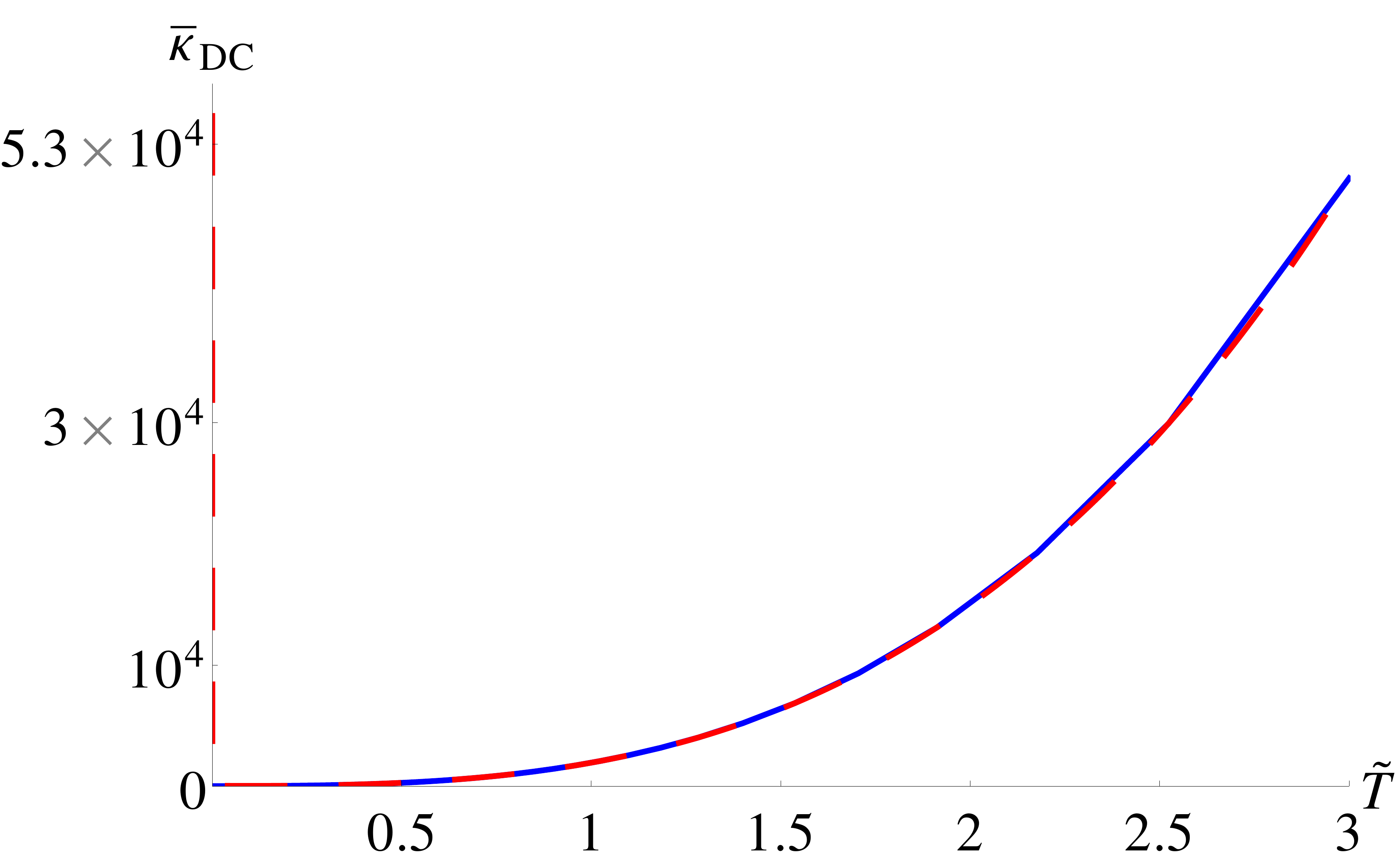}
\caption{Comparisons between the numerically computed (solid blue lines) thermo-electric conductivity $s_{DC}$ (left) and 
the numerically computed thermal conductivity $\bar{\kappa}_{DC}$ (right) with the hydrodynamic formul\ae\ 
\eqref{sart} and \eqref{kappaart} (red dashed lines) for $\beta=-1.04, \; \mu=1, \; L=1, \;$ and $\gamma=0.6$.}
\label{seebeckcomp}
\end{figure}

\begin{figure}[h!]
\centering
\includegraphics[width=7.9cm]{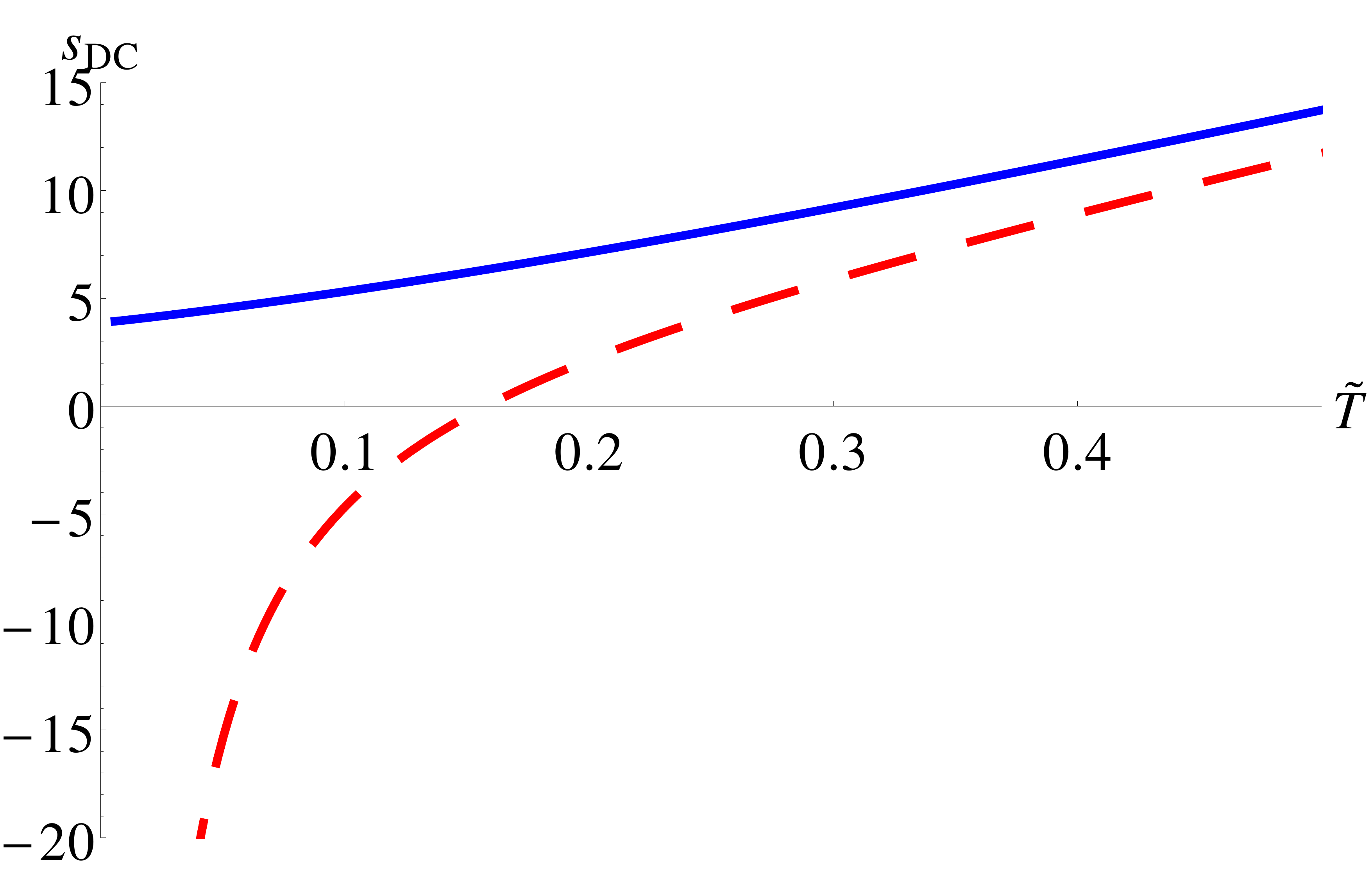}
\includegraphics[width=7.9cm]{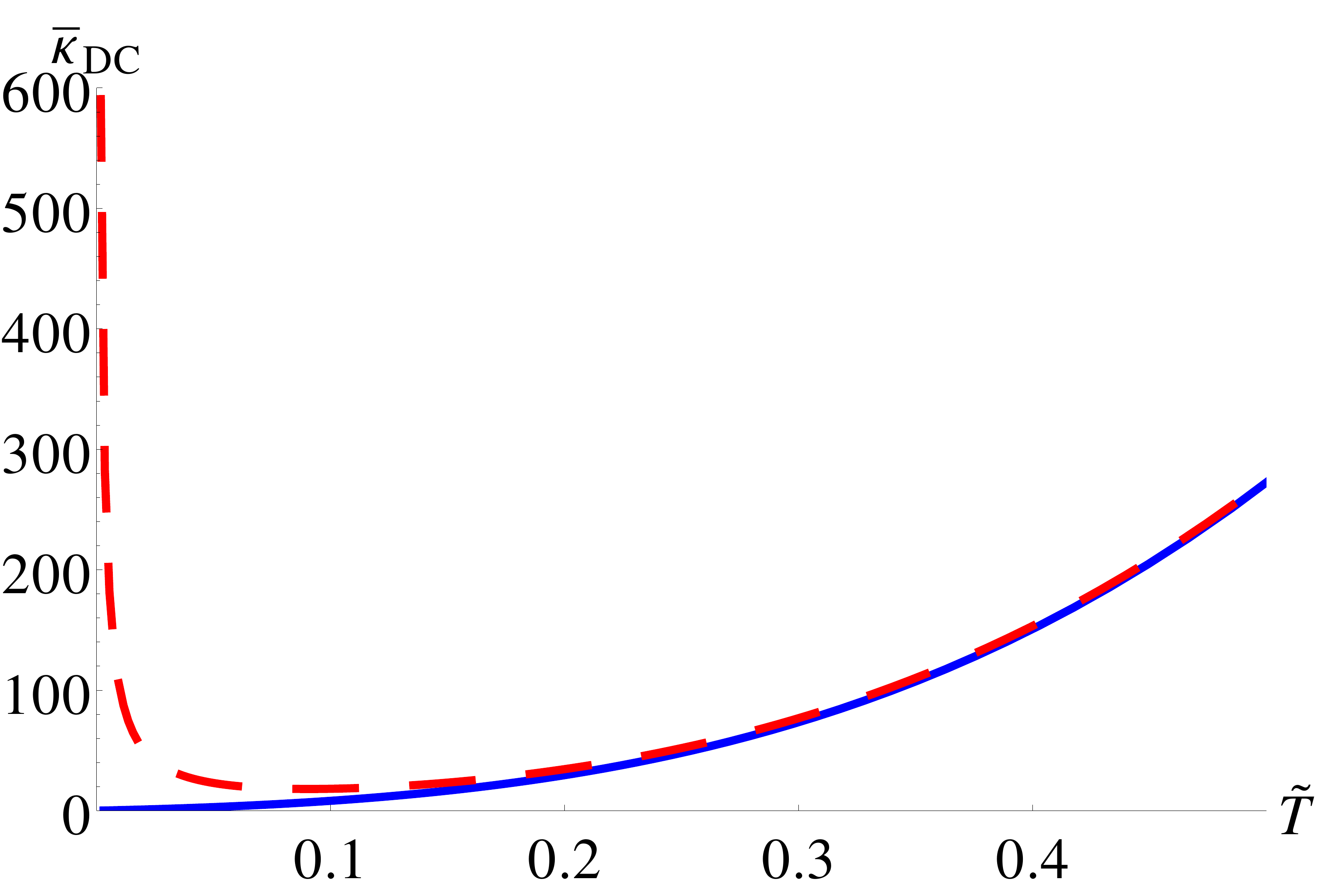}
\caption{A magnification in the low $\tilde{T}$ region of the comparisons between the numerically computed 
(solid blue lines) thermo-electric conductivity $s_{DC}$ (left) and the numerically computed thermal conductivity $\bar{\kappa}_{DC}$ 
(right) with the hydrodynamic formul\ae\ \eqref{sart} and \eqref{kappaart} (red dashed lines) for $\beta=-1.04, \; \mu=1, \; L=1, \;$ and $\gamma=0.6$.}
\label{kappacomp1}
\end{figure}

The comparison between the thermal conductivity \eqref{kappa} and the Seebeck coefficient \eqref{seebeck} computed numerically 
(blue solid lines) and the corresponding hydrodynamic formul\ae\ \eqref{sart}, \eqref{kappaart} (red dashed lines) are plotted in Figures \ref{seebeckcomp} and \ref{kappacomp1}. 
All the numerical computations whose results are shown in the plots are obtained taking $\mu=1$
and for a particular choice of the parameters $\beta, \; L, \; q$ and $\kappa_4$. Nevertheless, it is essential to mention that 
the various behaviors plotted are qualitatively the same for all the allowed values one could choose for this quantities%
\footnote{In particular, we recall that the scaling symmetry \eqref{scalingg} allows us to fix $\gamma$ and to vary the chemical potential or vice-versa.}.


From Figure \ref{seebeckcomp} emerges that, in the high $\tilde{T}$ region the transport coefficients 
computed numerically match exactly the hydrodynamic expectation \eqref{sigmaart}-\eqref{kappaart}. 
This confirms that, as proven in \cite{Davison:2013jba}, the massive gravity model under study has a hydrodynamic regime
that is well described by means of the modified conservation law \eqref{cons2}.
On the other hand, in the low-$\tilde{T}$ region (magnified in Figure \ref{kappacomp1})
the hydrodynamic description deviates from our numerical results.
% On the other hand, in the low-$\tilde{T}$ region (magnified in Figure \ref{kappacomp1})
% the hydrodynamic condition $\tau^{-1} \leq T$ is not satisfied and we find a disagreement between the 
% hydrodynamic description and our numerical results.
In particular note that the hydrodynamical plots coming from both \eqref{sart} and \eqref{kappaart} 
diverge as $\tilde{T} \rightarrow 0$; this clearly indicates the intrinsic limit of the hydrodynamic description at low $\tilde{T}$%
\footnote{Another sign of the hydrodynamic weakness at low $\tilde{T}$ emerges form the fact that
the Seebeck coefficient changes sign, which appears quite an unjustifiable feature within the model considered.}.
As we will further comment in Subsection \ref{beyond}, the Seebeck coefficient computed numerically approaches instead 
a constant value and the numerical thermal conductivity goes linearly to 0.

\subsection{Beyond the hydrodynamic regime}
\label{beyond}
\begin{figure}[h!]
\centering
\includegraphics[width=7.9cm]{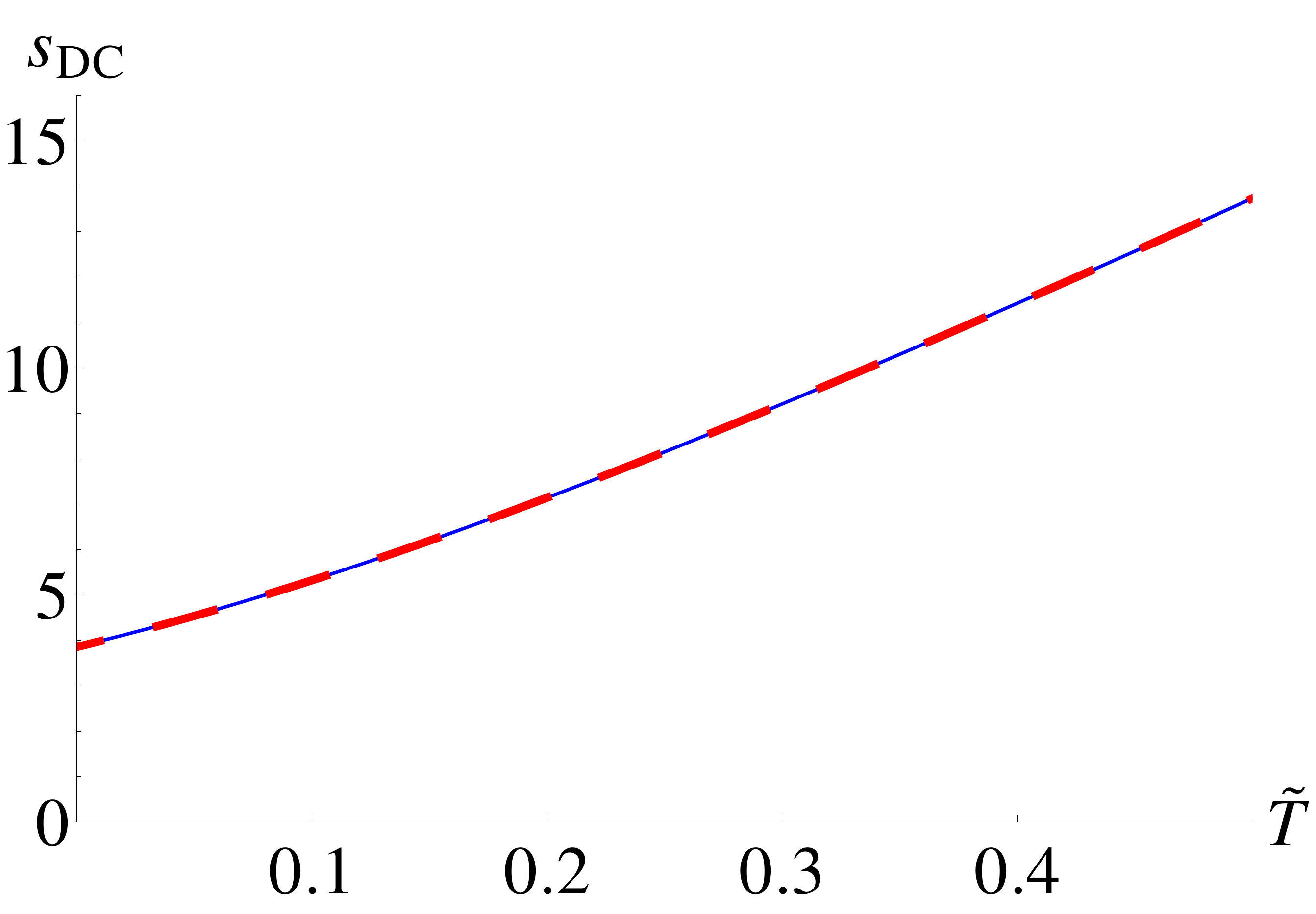}
\includegraphics[width=7.9cm]{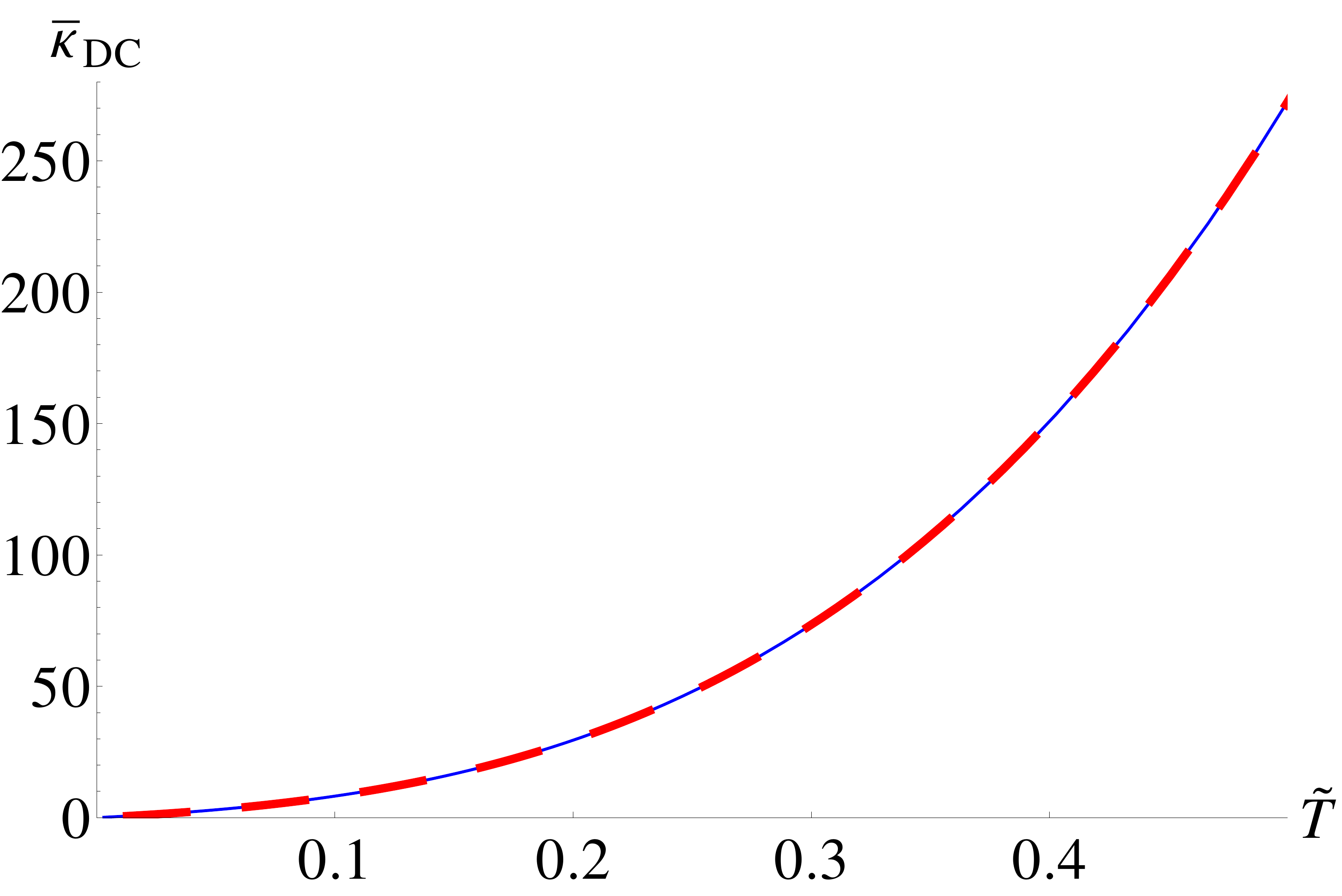}
\caption{Comparisons between the numerically computed (solid blue lines) thermo-electric conductivity $s_{DC}$ (left) and 
the numerically computed thermal conductivity $\bar{\kappa}_{DC}$ (right) with the formul\ae\ \eqref{seebecksach} 
and \eqref{kappasach} (red dashed lines) for $\beta=-1.04, \; \mu=1, \; L=1, \;$ and $\gamma=0.6$.}
\label{kappacomp}
\end{figure}
In the low-$\tilde{T}$ region the hydrodynamic approximation ceases to be valid.
Specifically, the hydrodynamic formul\ae\ do not agree with the Seebeck coefficient $s_{\text{DC}}$ 
and the thermal conductivity $\bar{\kappa}_{\text{DC}}$ obtained through our numerical computations.
Remarkably, as it is evident from Figure \ref{kappacomp}, in this region we find that our numerical 
results match exactly with the following expressions%
\footnote{This formul\ae\ has been confirmed by our analytical computation in \cite{Amoretti:2014mma}.}:
\begin{eqnarray}
\label{seebecksach}
s_{DC} & = &\frac{\mathcal{S} \rho}{\mathcal{E}+P}\tau  \ ,\\
\label{kappasach}
\bar{\kappa}_{DC} & = &\frac{\mathcal{S}^2 T}{\mathcal{E}+P}\tau \ .
\end{eqnarray}
The previous formul\ae\ allow us to be more quantitative and precise about the range of validity of the hydrodynamic regime in this model. 
Specifically, by comparing expressions \eqref{seebecksach} and \eqref{kappasach} with the hydrodynamic expressions \eqref{sart} and \eqref{kappaart} 
it is easy to see that hydrodynamics is a good approximation when the two following constraints are both satisfied%
\footnote{We thank the referee for suggesting us this precise comparison and formul\ae.}:
\begin{equation}\label{fc}
\beta \ll 2 \pi \rho q^2 T/ \mu, \qquad \beta \ll 2 \pi \mathcal{S} q^2 T^2/\mu^2 \ .
\end{equation} 
We remark that in this regime the model has many features in common with the behavior of the Fermi-liquid in the disorder dominated regime.
First of all, the Wiedemann-Franz ratio $\kappa_{DC}/(\sigma_{DC} T)$ (where $\kappa_{DC} = \bar{\kappa}_{DC}-s_{DC}^2T/\sigma_{DC}$) is constant in temperature even though its numerical value $L_0$,
\begin{equation}
L_0=\frac{2 \pi ^2 L^2 q^2 \left(\kappa_4 ^2 \mu ^2-2 \beta  L^2 q^2\right)}{3 \kappa_4 ^2 
\left(\kappa_4 ^2 \mu ^2-\beta  L^2 q^2\right)} \ ,
\end{equation} 
depends on $\beta$ and in general disagrees with the Fermi-liquid prediction $L_0^{(\text{Fl})} = \pi^2/3$ \footnote{For a recent discussion on the Wiedemann-Franz law in strongly correlated systems see \cite{Mahajan:2013cja}}.
Additionally, the thermal conductivity $\bar \kappa_{DC}$ goes linearly to zero with the temperature and is 
proportional to the heat capacity $C=T \frac{\partial \mathcal{S}}{\partial T}$, 
\begin{equation}
\bar \kappa_{DC}=-\frac{\sqrt{\frac{3}{2} \gamma ^2 \mu ^2 L^2 -3 \beta  L^4 }}{2 \beta  L^2 -2 \gamma ^2 \mu ^2} C +\mathcal{O}\left(T^2 \right) \ ,
\end{equation} 
while the electric conductivity $\sigma_{DC}$ is independent of the temperature, which constitutes 
another feature of the Fermi-liquid disorder-dominated regime.

The comparison of our model at low-$\tilde{T}$ with the disorder-dominated 
Fermi-liquid appears however to be not complete. In particular, the Mott law
describing the Fermi-liquid thermo-electric response
\begin{equation}
\label{mottlaw}
s_{\text{Mott}}=-\frac{\pi^2}{3 q^2}\ T\ \frac{\partial \log{\sigma_{DC}}}{\partial \mu}
\end{equation}
is not satisfied even qualitatively. In fact equation \eqref{mottlaw} yields a Seebeck coefficient 
which goes linearly to zero as $T \rightarrow 0$. On the contrary, as we have already noted, in the system at hand, $s_{DC}$ 
approaches a constant at $T=0$ and then grows linearly with the temperature:
\begin{equation}
s_{DC}=\frac{\sqrt{2} \pi  \mu  \left(2 \beta  L^2 -\gamma ^2 \mu ^2\right)}{\beta  q^2 \sqrt{3 \gamma ^2 \mu ^2 L^2 -6 \beta  L^4
   }}-\frac{4 \pi ^2 \mu  }{3 \beta  q^2}T+O\left(T^2\right)
\end{equation}
This is due to the fact that the entropy $\mathcal{S}$ is non-zero at $T=0$ (see \eqref{thermomassive} and \eqref{seebecksach}) . 

It is important to stress further that the formul\ae\ for the transport coefficients \eqref{seebecksach} and \eqref{kappasach} 
are exactly those computed in \cite{sachdev} for Dirac fermions with fermion-fermion interactions and a dilute density of charged 
impurities using the Boltzmann approach in the large-doping regime $\mu \gg T$, (formul\ae\ (6.4) and (6.5) of \cite{sachdev}). 
This fact, together with the many features which massive gravity has in common with the Fermi-liquid in this regime suggests that, 
at least in the large-doping region, a quasi-particle descriptions may be accurate. 
However, to prove the existence of a quasi-particle regime in the present model requires a systematic and careful analysis of the quasi-normal modes 
of the gravity solutions which we postpone to future investigations \cite{amoretti}.

\section{Conclusion and future prospect}
\label{conclu}

 We have throughly studied and characterized the thermo-electric transport 
of a simple holographic model featuring momentum dissipation in the boundary theory.
We regard the results obtained as interesting both from a purely theoretical perspective and
from a phenomenological standpoint. Regarding the former, we demonstrated the possibility 
of obtaining a physically consistent picture for the thermo-electric response of a gauge/gravity model 
possessing massive gravitons in the bulk. This feature leads to a breaking of some diffeomorphism
in the gravity model which therefore has a lower amount of symmetry. Therefore, performing the 
holographic renormalization of a massive gravity model, one must consider a larger set of possible counter-terms. 
The additional freedom proves crucial in obtaining a consistent phenomenological picture because 
the appropriate choice of finite counter-terms allows one to prevent the appearance of an unphysical 
dissipation-less heat transport mode at null frequency.

From a more phenomenologically-oriented viewpoint, it is tantalizing to observe the closeness between the transport properties
of the model at hand and the physics of the crossover between the quantum-critical to Fermi-liquid regimes
discussed for the graphene. The behavior of the model at hand in the limiting high and low temperature regions respectively
is in agreement with the non-holographic expectation of a hydrodynamic and quasi-particle regimes. 
On top of that, the holographic model allows one to study also intermediate regimes and offers the 
opportunity of having a complete setup interpolating the asymptotic regions.

A noteworthy fact is the possibility of the emergence, in our model, of a quasi-particle-like regime in the low-temperature 
region which is usually based on a standard Boltzmann description of quasi-particle degrees of freedom.
Such description is not immediately connected with a microscopic detail of the model; indeed
this quasi-particle regime arises in the deep IR (actually $\omega = 0$). At any rate, it is interesting to observe that a Fermi-liquid-like
physical picture can arise from a strongly coupled, momentum dissipating gauge/gravity model.
Of course, more investigation is needed in this respect. Both the assessment of this Fermi-liquid behavior
and its detailed dynamics call for further exploration (e.g. the study of probe fermions on the 
massive gravity charged black hole).

In the writing of this paper we became aware of related studies about the thermo-electric transport in holographic systems with momentum dissipation \cite{Donos:2014cya}. 
In \cite{Donos:2014cya} the momentum dissipation is realized by means of additional scalar fields within the Q-lattice framework. 
Remarkably, the analytic formul\ae\ for the thermo-electric transport coefficients found in \cite{Donos:2014cya} are compatible with those found by us 
in the contest of massive gravity. It would be interesting to further investigate the relation between this two results (see \cite{Amoretti:2014mma}).

One natural extension of the present analysis consists in studying the quasi-normal modes of the 
system in the whole temperature range. In other words, the extension of the study presented in \cite{Davison:2013jba}
to the ballistic and intermediate regimes as well. Although possibly technically demanding, such an analysis could shed light 
on the intimate nature of the holographic plasma and some statements regarding the quasi-particle nature of the
low-temperature physics could obtain conclusive evidence.

Another very promising direction for further work is represented by the inclusion of a magnetic field%
\footnote{A closely related analysis has been illustrated in \cite{Blake:2014yla}.}.
This not only allows one to study the mixed magnetic, electric and thermal transport, but could offer the 
possibility of studying other features which are based on experimental expectations. In particular, the presence 
of cyclotron modes which are intrinsically related to a collective nature of the quantum critical response.

\section{Acknowledgements}

A.A. and D.M. would like to address a particular thank to Richard Davison for many
discussions and key suggestions about many important points of the project.\\
D.M. thanks Andrea Mezzalira, Davide Forcella, Diego Redigolo, Subir Sachdev, David Vegh and Aristomenis Donos
for their comments and suggestions.\\
A. B. thanks the MIUR-FIRB2012 - Project HybridNanoDev (Grant
No. RBFR1236VV)
 and European Union FP7/2007-2013
under REA grant agreement no 630925 - COHEAT.

\appendix

\section{Transport matrix in the momentum dissipating case. Computational details}
\label{compumass}

Keeping into account the boundary expansions of the fluctuation fields \eqref{espUV}, 
the on-shell action \eqref{boundaryac} reads
\begin{multline}
\label{acci1}
S_{\text{tot}} = V \int \frac{d \omega}{2 \pi} 
\Big[\frac{1}{2q^2L} a_x^{(0)} a_{x}^{(1)} 
- \frac{\rho}{q^2} \frac{\beta + \omega^2}{2 \beta + \omega^2} h_{tx}^{(0)} a_x^{(0)}\\
- \frac{3 \beta}{2\kappa^2_4 L (2\beta + \omega^2)} h_{tx}^{(0)} h_{tx}^{(1)} 
- (1-a) \frac{\mathcal{E}}{4} h_{tx}^{(0)} h_{tx}^{(0)} \Big] \ +  \left(\omega \leftrightarrow -\omega \right) \ ,
\end{multline}
where the arguments of the first and second fluctuation field in each term are respectively $-\omega$ and $\omega$.
In order to simplify the notation, we introduce gothic letters to indicate the coefficients in the quadratic action:
\begin{equation}
\label{acci2}
S_{\text{tot}} = V \int \frac{d \omega}{2 \pi} 
\Big[ \mathfrak{A}\ a_x^{(0)} a_{x}^{(1)} 
+ \mathfrak{B}\ h_{tx}^{(0)} a_x^{(0)}
+ \mathfrak{C}\ h_{tx}^{(0)} h_{tx}^{(1)} 
+ \frac{\mathfrak{D}}{2}\ h_{tx}^{(0)} h_{tx}^{(0)} \Big] \ + \left( \omega \leftrightarrow -\omega \right) \ ,
\end{equation}
where the correspondence between gothic letters and coefficients is easily understood by comparing \eqref{acci2} with \eqref{acci1}.

The relation between the derivatives with respect to the physical quantities and those 
with respect to the sources of the bulk fields is given in \eqref{derivs}.
We remind the reader that the sources $h_{tx}^{(0)}$ and $a_x^{(0)}$ are independent 
and the derivative with respect to one of them is taken putting the other to zero 
(this fact is understood throughout our formul\ae).
The off-diagonal term in the transport matrix is due to the mixed second order derivative.  Exploiting linearity we obtain
\begin{equation}
 \begin{split}
 \frac{\delta S}{\delta E} &= -\left( \frac{i}{\omega}\right)\frac{\delta S}{\delta a_x^{(0)}} \\
  &=-\left( \frac{i}{\omega}\right) \left( \mathfrak{A}\ a_{x}^{(1)} + \mathfrak{B}\ h_{tx}^{(0)} + \mathfrak{C}\ h_{tx}^{(0)} \frac{\delta h_{tx}^{(1)}}{\delta a_x^{(0)}} + \mathfrak{A}\ a_x^{(0)} \frac{\delta a_{x}^{(1)}}{\delta a_x^{(0)}} \right) \ ,
 \end{split}
\end{equation}
\begin{equation}
 \begin{split}
  - T\frac{\delta^2 S}{\delta \nabla T \delta E} &=  -\left( \frac{i}{\omega}\right) \left(\frac{\delta}{\delta h_{tx}^{(0)}} - \mu \frac{\delta}{\delta a_x^{(0)}}\right) \frac{\delta S}{\delta E}\\
  &= \left( \frac{i}{\omega}\right)^2 \left( \mathfrak{A}\ \frac{\delta a_{x}^{(1)}}{\delta h_{tx}^{(0)}} + \mathfrak{B} + \mathfrak{C}\ \frac{\delta h_{tx}^{(1)}}{\delta a_x^{(0)}} - 2 \mu \mathfrak{A}\ \frac{\delta a_{x}^{(1)}}{\delta a_x^{(0)}} \right) \ .
 \end{split}
\end{equation}
Since we are dealing with a system preserving time-reversal symmetry, 
due to Onsager's argument the transport matrix must be symmetric.
To check the symmetrical character of the transport matrix offers a useful check
of the correctness of the computations (which is slightly delicate due to the non-trivial 
relation between the physical and the bulk fields). On a technical ground, we need to verify that the 
functional derivatives commute, namely
\begin{equation}
 \begin{split}
 -T\frac{\delta S}{\delta \nabla T} =&  -\left( \frac{i}{\omega}\right) \left(\frac{\delta}{\delta h_{tx}^{(0)}} - \mu \frac{\delta}{\delta a_x^{(0)}}\right) S\\
 =&  -\left( \frac{i}{\omega}\right) \Big( \mathfrak{A}\ a_x^{(0)} \frac{\delta a_{x}^{(1)}}{\delta h_{tx}^{(0)}} + \mathfrak{B}\ a_x^{(0)} + 2 \mathfrak{D}\ h_{tx}^{(0)} + \mathfrak{C}\ h_{tx}^{(1)}\\
 &+ \mathfrak{C}\ h_{tx}^{(0)} \frac{\delta h_{tx}^{(1)}}{\delta h_{tx}^{(0)}} - \mu \mathfrak{A}\ a_{x}^{(1)} - \mu \mathfrak{A}\ a_x^{(0)} \frac{\delta a_{x}^{(1)}}{\delta a_x^{(0)}}
 - \mu \mathfrak{B}\ h_{tx}^{(0)} - \mu \mathfrak{C}\ h_{tx}^{(0)} \frac{\delta h_{tx}^{(1)}}{\delta a_x^{(0)}} \Big)
 \end{split}
\end{equation}
\begin{equation}
 \begin{split}
 -T \frac{\delta^2 S}{\delta E \delta \nabla T} =&  \left( \frac{i}{\omega}\right)^2 \left( \mathfrak{A}\ \frac{\delta a_{x}^{(1)}}{\delta h_{tx}^{(0)}}
 + \mathfrak{B} + \mathfrak{C}\ \frac{\delta h_{tx}^{(1)}}{\delta a_x^{(0)}} - 2 \mu \mathfrak{A}\ \frac{\delta a_{x}^{(1)}}{\delta a_x^{(0)}} \right) \ .
 \end{split}
\end{equation}
We have the right commutation between the derivatives and taking stock of the preceding computations, we have
\begin{equation}
 \begin{split}
 -T \frac{\delta^2 S}{\delta E \delta \nabla T} =& - T\frac{\delta^2 S}{\delta \nabla T \delta E}
 =  \left( \frac{i}{\omega}\right)^2 \left(\mathfrak{A}\ \frac{\delta a_{x}^{(1)}}{\delta h_{tx}^{(0)}}
 + \mathfrak{B} + \mathfrak{C}\ \frac{\delta h_{tx}^{(1)}}{\delta a_x^{(0)}} - 2 \mu \mathfrak{A}\ \frac{\delta a_{x}^{(1)}}{\delta a_x^{(0)}} \right) \ .
 \end{split}
\end{equation}
Repeating the same steps for the diagonal entries of the transport matrix \eqref{traspo}, we obtain:
\begin{equation}
 \frac{\delta^2 S}{\delta E^2} = \frac{\delta^2 S}{(\delta a_x^{(0)})^2} = 2 \left( \frac{i}{\omega}\right)^2\, \mathfrak{A}\ \frac{\delta a_{x}^{(1)}}{\delta a_x^{(0)}} \ ,
\end{equation}
\begin{equation}
 \begin{split}
 T^2\frac{\delta^2 S}{\delta \nabla T^2} 
 = 2 \left( \frac{i}{\omega}\right)^2 \left[ \mathfrak{D}\ + \mathfrak{C}\ \frac{\delta h_{tx}^{(1)}}{\delta h_{tx}^{(0)}}
  - \mu \mathfrak{A} \frac{\delta a_{x}^{(1)}}{\delta h_{tx}^{(0)}} - \mu \mathfrak{B}
  - \mu \mathfrak{C}\ \frac{\delta h_{tx}^{(1)}}{\delta a_x^{(0)}}
  + \mu^2 \mathfrak{A} \frac{\delta a_{x}^{(1)}}{\delta a_x^{(0)}}\right] \ .
 \end{split}
\end{equation}
In conclusion, due to the linearity requirement, the transport coefficients are:
\begin{equation}
\sigma=-\left. \frac{1}{q^2L}\frac{i}{\omega} \frac{ \delta a_{x}^{(1)}}{ \delta a_x^{(0)}}\right|_{h_{tx}^{(0)}=0} \ ,
\end{equation}
\begin{multline}
\bar{\kappa} = -2\frac{i}{T \omega} \Big[(a-1)\frac{\mathcal{E}}{2} - \frac{3 \beta}{2\kappa_4^2 L(2\beta + \omega^2)} \left. \frac{ \delta h_{tx}^{(1)}}{ \delta h_{tx}^{(0)}} \right|_{a_x^{(0)}=0}
  -  \frac{\mu}{2q^2L} \left. \frac{ \delta a_{x}^{(1)}}{ \delta h_{tx}^{(0)}} \right|_{a_x^{(0)}=0} +\\  \frac{\mu^2}{z_h q^2} \frac{\beta + \omega^2}{2 \beta + \omega^2}
  +  \frac{3 \mu \beta}{2\kappa_4^2 L(2\beta + \omega^2)} \left. \frac{ \delta h_{tx}^{(1)}}{ \delta a_x^{(0)}} \right|_{h_{tx}^{(0)}=0}
  +  \frac{\mu^2}{2q^2L} \left. \frac{ \delta a_{x}^{(1)}}{ \delta a_x^{(0)}} \right|_{h_{tx}^{(0)}=0}  \Big]
\end{multline}
\begin{multline}
s =-\frac{i}{T \omega}\Big[ \frac{1}{2q^2L} \left.\frac{\delta a_{x}^{(1)}}{\delta h_{tx}^{(0)}}\right|_{a_x^{(0)}=0}
 - \frac{\mu}{z_h q^2} \frac{\beta + \omega^2}{2 \beta + \omega^2}\\ - \frac{3 \beta}{2\kappa_4^2 L(2\beta + \omega^2)} \left.\frac{ \delta h_{tx}^{(1)}}{\delta a_x^{(0)}}\right|_{h_{tx}^{(0)}=0} 
 -   \frac{\mu}{q^2L} \left.\frac{\delta a_{x}^{(1)}}{\delta a_x^{(0)}}\right|_{h_{tx}^{(0)}=0} \Big] \ .
\end{multline}

\end{document}